\documentclass{bmcart}

\usepackage{url,hyperref,lineno,microtype,subcaption}
\usepackage{amsthm,amsmath}
\usepackage[utf8]{inputenc} 

\usepackage{graphicx}
\usepackage{dcolumn}
\usepackage{bm}
\usepackage{mathdots}
\usepackage{siunitx}
\usepackage{tikzorbital}
\usepackage{braket}
\usepackage{csvsimple}
\newcommand{\estimates}{\overset{\scriptscriptstyle\wedge}{=}}

\usepackage{pgfplots}
\pgfplotsset{my style/.append style={xlabel={$R$(\AA)}, ylabel={LiH Energy (Hartree)}}}

\pgfplotsset{H2 style/.append style={xlabel={$R$(\AA)}, ylabel={H$_2$ Energy (Hartree)}}}

\startlocaldefs
\endlocaldefs

\begin{document}

\begin{frontmatter}

\begin{fmbox}
\dochead{Research}

\title{Periodic Plane-Wave Electronic Structure Calculations on Quantum Computers}

\author[id=m1,
   addressref={aff1},                   
   corref={aff1},
   email={duo.song@pnnl.gov}   
]{\inits{DS}\fnm{Duo} \snm{Song}}
\author[
   addressref={aff1},                   
   email={nicholas.bauman@pnnl.gov}   
]{\inits{NB}\fnm{Nicholas P} \snm{Bauman}}
\author[
   addressref={aff2},
   email={guenp@microsoft.com}
]{\inits{GP}\fnm{Guen} \snm{Prawiroatmodjo}}
\author[
   addressref={aff1},                   
   email={peng398@pnnl.gov}   
]{\inits{BP}\fnm{Bo} \snm{Peng}}
\author[
   addressref={aff2},
   email={casandra.granade@microsoft.com}
]{\inits{CR}\fnm{Cassandra} \snm{Granade}}
\author[
   addressref={aff1},
   email={kevin.rosso@pnnl.gov}
]{\inits{KMR}\fnm{Kevin M} \snm{Rosso}}
\author[
   addressref={aff2},
   email={guanghao.lowe@microsoft.com}
]{\inits{GHL}\fnm{Guang Hao} \snm{Low}}
\author[
   addressref={aff2},
   email={martinro@microsoft.com}
]{\inits{MR}\fnm{Martin} \snm{Roetteler}}
\author[
   addressref={aff1},
   email={karol.kowalski@pnnl.gov}
]{\inits{KK}\fnm{Karol} \snm{Kowalski}}

\author[id=n1,
   addressref={aff1},                   
   corref={aff1},                       
   email={eric.bylaska@pnnl.gov}   
]{\inits{EJB}\fnm{Eric J} \snm{Bylaska}}

\address[id=aff1]{
  \orgname{Fundamental Sciences Division, Pacific Northwest National Laboratory}, 
  \city{Richland},                              
  \state{WA}
  \postcode{99354}
  \cny{USA}                                    
}

\address[id=aff2]{
  \orgname{Microsoft Quantum, Microsoft},     
  \city{Redmond},                             
  \state{WA}
  \postcode{98052}
  \cny{USA}                                    
}

\begin{artnotes}
\note[id=m1,id=n1]{Corresponding Authors} 
\end{artnotes}

\end{fmbox}

\begin{abstractbox}
\begin{abstract}
A procedure for defining virtual spaces, and the periodic one-electron and two-electron integrals, for plane-wave second quantized Hamiltonians has been developed, and it was validated using full configuration interaction (FCI) calculations, as well as executions of variational quantum eigensolver (VQE) circuits on Quantinuum's ion trap quantum computers accessed through Microsoft's Azure Quantum service. This work is an extension to periodic systems of a new class of algorithms in which the virtual spaces were generated by optimizing orbitals from small pairwise CI Hamiltonians, which we term as correlation optimized virtual orbitals with the abbreviation COVOs. In this extension, the integration of the first Brillouin zone is automatically incorporated into the two-electron integrals. With these procedures, we have been able to derive virtual spaces, containing only a few orbitals, that were able to capture a significant amount of correlation. The focus in this manuscript is on comparing the simulations of small molecules calculated with plane-wave basis sets with large periodic unit cells at the $\Gamma$-point, including images, to results for plane-wave basis sets with aperiodic unit cells. The results for this approach were promising, as we were able to obtain good agreement between periodic and aperiodic results for an LiH molecule. Calculations performed on the Quantinuum H1-1 quantum computer produced surprisingly good energies, in which the error mitigation played a small role in the quantum hardware calculations and the (noisy) quantum simulator results. Using a modest number of circuit runs (500 shots), we reproduced the FCI values for the 1 COVO Hamiltonian with an error of 11 milliHartree, which is expected to improve with a larger number of circuit runs.
\end{abstract}
                             

\end{abstractbox}

\end{frontmatter}

\section{Introduction}
\label{introduction}

With the arrival of quantum computers, researchers are actively developing new algorithms  to carry out quantum chemistry calculations on these platforms, in particular for calculations containing strong electron-electron correlations (aka high-level quantum chemistry methods).  This is because it is anticipated that quantum computers with 50-100 qubits will eventually surpass classical digital computers for these types of calculations~\cite{preskill2018quantum}. However, in order for quantum computing to reach its full potential, there are hardware and software challenges that need to be addressed before it can become a viable replacement~\cite{wasielewski2020exploiting} for existing high-performance classical computers and the associated cutting-edge parallel software that have been developed in the last two decades.

Most high-level quantum chemistry methods in use today (e.g., full configuration interaction (FCI)~\cite{szabo2012modern,ross1952calculations,gan2006lowest,mcardle2020quantum,tubman2020modern,sugisaki2021quantum,kawashima2021optimizing}, coupled cluster (CC)~\cite{coester58_421,coester60_477,cizek66_4256,paldus72_50,mukherjee1975correlation,purvis82_1910,bishop1987coupled,paldus07,crawford2000introduction,bartlett_rmp,peng2021coupled}, and Green's function (GF)~\cite{green1854essay,green2014essay,feynman1949theory,feynman1958Space,schwinger1959Theory,baym1961conservation,peng2021coupled} approaches) are based on second-quantized Hamiltonians, which are written in terms of the creation and annihilation operators of the Fermion orbitals along with the one-electron and two-electron integrals for the system.  In principle, this formulation is exact, however, conventional computing methods are restricted in their accuracy due to the prohibitive computational cost for exact modeling of the exponentially growing wavefunction from the basis set that is introduced.  As a result, these basis sets are typically highly engineered.  One of the first, and still popular, class of basis sets used in quantum chemistry methods are atomic-like orbitals or the linear combination of atomic orbitals (LCAO) basis set. Pioneered by J. Lennard-Jones~\cite{lennard1929electronic}, L. Pauling~\cite{pauling1931nature}, and J.C. Slater~\cite{slater1930atomic}, the atomic orbitals are generated by carrying out an atom calculation for each kind of atom in the system; guided by the heuristic that says the electronic states of a molecule or solid can be thought of as a superposition of atomic orbitals.  For high-level methods, a popular basis set is the Dunning correlation consistent basis set~\cite{dunning1977gaussian,dunning-cc,Prascher2011}, in which the atomic orbitals are optimized at the configuration interaction singles and doubles (CISD) level of theory~\cite{handy1980multi}.  While the size of this intuitive, optimistically a priori, class of basis set is small compared to modern style basis sets that are more complete, e.g., plane-wave basis sets, it still needs to contain enough atomic orbitals to produce a truly accurate result.

Another challenge is calculating the two-electron integrals for condensed phase systems,  since one typically wants to use periodic boundary conditions to carry out the simulation.  While this is natural for plane-wave DFT methods~\cite{bylaska2011large,bylaska2011parallel,bylaska2017plane,pickett1989electronic,ihm1979momentum,car1985unified,payne1992iterative,remler1990molecular,kresse1996efficient,marx2000modern,marx2009ab,martin2004electronic,ValievBylaskaWeare2002,bylaska2009hard,chen2016first,gygi2008architecture,bylaska2002parallel,peng2021coupled} with low levels of theory, it is significantly more complicated to calculate exact exchange~\cite{gygi1985exact,gygi1986self,gygi1989quasiparticle,bylaska2011parallel,bylaska2017plane,bylaska2020filon,chawla1998exact,sorouri2006accurate,marsman2008hybrid,gorling1996exact,distasio2014individual,peng2021coupled} and the other two-electron integrals~\cite{bylaska2018corresponding} with periodic boundary conditions, as it requires special integration strategies to handle the integration of the Brillouin zone.   At first glance, periodic many-body calculations would appear to be intractable because the expansion of one-electron orbitals in terms of Bloch states leads to a large number of orbitals describing the first Brillouin zone~\cite{bylaska2020filon},
\begin{equation*}
   \psi_{\sigma,n\mathbf{k}}(\mathbf{r}) = \frac{e^{i\mathbf{k} \cdot \mathbf{r}}}{\sqrt{\Omega}} \sum_{\mathbf{G}} \psi_{\sigma,n\mathbf{k}}(\mathbf{G}) e^{i \mathbf{G} \cdot \mathbf{r}},
\end{equation*}
where $\psi_{\sigma,n\mathbf{k}}(\mathbf{G})$ are the expansion coefficients,  $\Omega$ is the volume of the primitive cell ($\Omega=[\mathbf{a}_1,\mathbf{a}_2,\mathbf{a}_3]= \mathbf{a}_1\cdot (\mathbf{a}_2 \times \mathbf{a}_3)$), $\mathbf{r}$ is the position in real space, $\mathbf{G} $  are the reciprocal lattice vectors, $\sigma$ and $n$ are the spin and orbitals indexes, 
and $\mathbf{k}$ is a vector in the first Brillouin zone~\cite{kittel2005introduction,ashcroft1976introduction}.
Simple approximations to the integration over the Brillouin zone in the exact exchange and other two-electron integrals lead to very inaccurate results, e.g. a straightforward $\Gamma$-point approximated calculation results in the two-electron integrals being infinite~\cite{bylaska2018corresponding,bylaska2020filon}.  

To overcome these limitations, we have recently developed new methods for generating optimized orbital basis sets, called COVOs~\cite{bylaska2021quantum}.   This method is different from other plane-wave derived optimized orbital basis sets~\cite{shirley1996optimal,prendergast2009bloch,chen2011electronic} in that it is based on optimizing small select CI problems rather than fitting one-electron eigenvalue spectra and band structures. In this work, the COVOs method is extended to periodic systems at the $\Gamma$-point using the recently developed Filon integration strategy~\cite{bylaska2020filon} for calculating exact exchange energies and two-electron periodic integrals in electron transfer calculations~\cite{bylaska2018corresponding,bylaska2020electron,simonnin2021modeling}, in which the integration of the first Brillouin zone is automatically incorporated. 

In addition, present quantum devices are plagued by short coherence times and vulnerability to environment interference, i.e., noise. Although quantum algorithms such as quantum phase estimation can calculate molecular energies with proved exactness, these are not yet viable to run on near-term intermediate scale (NISQ) devices~\cite{preskill2018quantum, reiher2017}. Therefore, it is desirable to limit the operation of quantum processors to a complementary concerted execution with classical counterparts, whereby each of these components is only in charge of those tasks for which it is more suitable. This has materialized into the development of Variational Quantum Algorithms (VQA) ~\cite{peruzzo2014variational,cerezo2021}. Briefly, this class of algorithms strives to find the lowest eigenvalue of a given observable by assuming the associated quantum state can be accurately represented by a trial wave function and whose parameters are varied according to the Rayleigh-Ritz method (variational principle), with these parameters being updated by the classical computer. 

The paper is organized as follows. In section~\ref{sec:planewaveH}, a brief description of the second-quantized Hamiltonian and one-electron and two-electron integrals with periodic boundary conditions is given, followed in section~\ref{sec:periodic_covos} in which a new class of algorithm for generating a virtual space in which the orbitals are generated by minimizing small pairwise CI Hamiltonians. A complete set of equations for implementing these optimizations is given in subsections~\ref{sec:one-electron}-\ref{sec:two-electron-matrix}.
Using this new type of virtual space, CI calculations up to 18 virtual orbitals for the ground state energy curve of the LiH molecule in a periodic box are presented in section~\ref{sec:LiHresults}.  LiH is a commonly used test case in quantum computing~\cite{kandala2017,corcoles2019challenges,low2019q}. In section~\ref{sec:LiHQCresults}, results from quantum computing simulations using variational quantum eigensolver (VQE) quantum computing algorithms are presented, and lastly, the conclusions are given in section~\ref{sec:conclusions}.

\section{Pseudopotential Plane-Wave Second-Quantized Hamiltonian}
\label{sec:planewaveH}
The non-relativistic electronic Schr{\"o}dinger eigenvalue equation of quantum chemistry can be written as
\begin{equation}
\label{eqn:schrodinger}
H \ket{\Psi({\bf x}_1,{\bf x}_2,...,{\bf x}_{N_e})} = E \ket{\Psi({\bf x}_1,{\bf x}_2,...,{\bf x}_{N_e})}
\end{equation}
where $H$ is the electronic structure Hamiltonian under the Born--Oppenheimer approximation, and $\ket{\Psi({\bf x}_1,{\bf x}_2,...,{\bf x}_{N_e})}$ is the quantum mechanical wavefunction that is a function of the spatial and spin coordinates of the $N_e$ electrons, ${\bf x}_i = ({\bf r}_i, \sigma_i)$.  When solving this equation the Pauli exclusion principle constraint of particle exchange must be enforced, in which the wavefunction changes sign when the coordinates of two particles, ${\bf x}_i$ and ${\bf x}_j$, are interchanged, i.e. 
\begin{eqnarray}
\label{eqn:antisymmetry}
    && \ket{\Psi({\bf x}_1,{\bf x}_2,...{\bf x}_i,...{\bf x}_j,...,{\bf x}_{N_e})} \nonumber \\
    &=&
    -\ket{\Psi({\bf x}_1,{\bf x}_2,...{\bf x}_j,...{\bf x}_i,...,{\bf x}_{N_e})}.
\end{eqnarray}
 
For the Born--Oppenheimer Hamiltonian, the interaction between the electrons and nuclei are described by the proper potentials $\frac{Ze}{|{\bf r}_i - {\bf R}_A|}$, which for plane-wave solvers can cause trouble with convergence because of the singular behavior at $|{\bf r} - {\bf R}_A|$.  A standard way to remove this issue in plane-wave calculations is to replace these singular potentials by pseudopotentials.  By making this replacement, the Hamiltonian, $H$, in Eq.~\ref{eqn:schrodinger} can be written as
\begin{eqnarray*}
    H &=& -\frac{1}{2}\sum_{i=1}^{N_e} \nabla_i^2 \nonumber \\ 
    &+& \sum_{i=1}^{N_e} \sum_{A=1}^{N_A} \left(V^{(A)}_{local}(|{\bf r}_i - {\bf R}_A|) + \sum_{lm} \hat{V}^{(A),lm}_{NL} \right) \nonumber \\
    &+&  \sum_{i=1}^{N_e} \sum_{j>i}^{N_e} \frac{1}{|{\bf r}_i - {\bf r}_j|}
\end{eqnarray*}
where the first term is the kinetic energy operator, the second term contains the local and non-local pseudpotentials, $V^{(A)}_{local}$ and $\hat{V}^{(A),lm}_{NL}$, that represent the electron-ion interactions, and the last term is the electron-electron repulsion.

Instead of writing the many-body Hamiltonian in the traditional Schr{\"o}dinger form, as in the equations above, it is more common today to write it in an alternative representation, known as the second-quantization form.  In this form,  single particle (electron) creation $a_p^\dagger \ket{0} = \ket{1}$ and annihilation $a_p\ket{1} = \ket{0}$ operators are introduced, where the occupation of a specified state $p$ is defined as $\ket{1}$ and $\ket{0}$ for the occupied and unoccupied orbitals respectively. The second-quantized Hamiltonian is written as~\cite{bylaska2021quantum}
    \begin{equation} \label{eqn:secondH}
        H = \sum^{N_\text{basis}}_{p=1} \sum^{N_\text{basis}}_{q=1} h_{pq}a_p^\dagger a_q + \frac{1}{2}\sum_{pqrs}
        h_{pqrs}
        a_p^\dagger a_r^\dagger a_s a_q,
    \end{equation}
    \begin{eqnarray*}
        h_{pq} &=& \int d{\bf x} \psi_p^*({\bf x})\left( -\frac{1}{2} \nabla^2
         \right)\psi_q({\bf x}) \\
        +&&\int d{\bf x} \psi_p^*({\bf x})\left(\sum_{A=1}^{N_A} 
        \left(V^{(A)}_\text{local}(|{\bf r} -{\bf R}_A|) + \sum_{lm} \hat{V}^{(A),lm}_\text{NL} \right)
        \right)\psi_q({\bf x}) \\
        h_{pqrs} &=& \iint d{\bf x}_1 d{\bf x}_2 \psi_p^*({\bf x}_1)\psi_r^*({\bf x}_2)\frac{1}{|{\bf r}_1 - {\bf r}_2|}\psi_s({\bf x}_2) \psi_q({\bf x}_1)
    \end{eqnarray*}
where $\psi_p\left({\bf x}\right)$ represent one-electron spin-orbital basis.
A nice feature about this form of the Hamiltonian is that the antisymmetry of wavefunction requirement as given in Eq.~\ref{eqn:antisymmetry} is automatically enforced through the standard Fermionic anti-commutation relations $\{a_p,a_q^\dagger\}=\delta_{pq}$ and $\{a_p,a_q\}=\{a_p^\dagger,a_q^\dagger\}=0$. 

In this formulation,  the choice of the one-electron spin-orbital basis is nebulous and requires some care in its choosing in order to obtain accurate results with this type of Hamiltonian.  Typically, in quantum chemistry one uses the filled and virtual orbitals from a Hartree--Fock calculation.  For methods that utilize linear combinations of atomic orbitals (LCAO) as the basis, the size of the basis set and subsequently generated Hartree--Fock orbitals is fairly small.  However, for plane-wave solvers, and other grid based solvers, the size of the basis set is very large and the number of the one- and two-electron integrals in Eq.~\ref{eqn:secondH} will become prohibitive if all possible Hartree--Fock orbitals are used.  

We note the formulae for the one-electron and two-electron integrals in subsections \ref{subsec:1-electron-pspw}, \ref{subsec:2-electron-pspw} and \ref{subsection:ewald} are given in terms of the spatial orbitals rather than spin orbitals.  The spin functions $\alpha$ and $\beta$ are integrated out in the standard way, to involve only spatial functions and integrals~\cite{szabo2012modern}.  Many of the periodic forms presented in the following sections are written in terms of Fourier space using periodic plane-wave basis sets, rather than real space.  Descriptions of the plane-wave methods used in this work can be found in the following references~\cite{bylaska2011large,bylaska2011parallel,bylaska2017plane,bylaska2018corresponding,bylaska2020filon,pickett1989electronic,ihm1979momentum,car1985unified,payne1992iterative,remler1990molecular,kresse1996efficient,marx2000modern,martin2004electronic,ValievBylaskaWeare2002,chen2016first}.

\subsection{Periodic One-Electron Integrals using the Pseudopotential Plane-Wave Method}
\label{subsec:1-electron-pspw}
The one-electron integrals in the pseudopotential plane-wave method can be written as a sum of the kinetic, local pseudopotential and non-local pseudopotential energies~\cite{bylaska2017plane}.
\begin{equation}
h_{pq} = E^{pq}_\text{kinetic} + E^{pq}_\text{local}  + E^{pq}_\text{non-local} 
\label{eqn:1e}
\end{equation}

The kinetic energy can be written as
\begin{equation*}
E^{pq}_\text{kinetic} = \frac{1}{2}\sum_{\mathbf{G}} \mathbf{G}^2 \psi_p^{*}(\mathbf{G}) \psi_q(\mathbf{G}) 
\end{equation*}
where $\psi_p(\mathbf{G})$ and $\psi_q(\mathbf{G})$ molecular orbitals in Fourier space.  The local pseudopotential energy can be evaluated as
\begin{equation*}
E^{pq}_\text{local} = \sum_I \int_{\Omega} V_\text{local}^{I}(\mathbf{r}) \psi^{*}_p(\mathbf{r}) \psi_q(\mathbf{r}) d\mathbf{r} = \sum_{I,\mathbf{G}} V_\text{local}^I(\mathbf{G}) \rho_{pq}(\mathbf{G}) 
\end{equation*}
where the valence overlap electron density in reciprocal space $\rho_{pq}(\mathbf{G})$ is obtained from taking the fast Fourier transform of its real-space representation, $\rho_{pq}(\mathbf{r}) = \psi^{*}_p(\mathbf{r}) \psi_q(\mathbf{r})$. The local potential is defined to be periodic and is represented as a sum of piecewise functions on the Bravais lattice by
\begin{equation*}
V_\text{local}^I(\mathbf{r}) = \sum_{\mathbf{L}} V_\text{local}^I(|\mathbf{r}-\mathbf{R}_I - \mathbf{L}|)
\end{equation*}
where $\mathbf{R}_I$ is the location of the atom, I, in the unit cell, $\mathbf{L}$ is a Bravais lattice vector, and $V_\text{local}^{I}(r)$ is a radial local pseudopotential for the atom obtained from a Kleinman-Bylander expansion of a norm-conserving pseudopotential~\cite{kleinman1982efficacious,hamann1989generalized}. The local pseudopotential in reciprocal space can be generated by using an ($l=0$) spherical Bessel transform.
\begin{equation}\label{eqn:V_loc}
V_\text{local}^{I}(\mathbf{G}) = \frac{4\pi}{\sqrt{\Omega}} e^{ \mathit{i} \mathbf{G} \cdot \mathbf{R}_I} \int_{0}^{\infty} V_\text{local}^{I}(r) j_0(|\mathbf{G}|r) r^2 dr
\end{equation}
where $j_0(x) = \frac{sin(x)}{x}$ is the $l=0$ spherical Bessel function of the first kind. The non-local pseudopotential energy can be evaluated as
\begin{equation}\label{eqn:E_NL}
E^{pq}_\text{non-local} = \sum_I \sum_{lm} \sum_{n,n'}  \Bigg[\sum_{\mathbf{G}} \psi_p^{*}(\mathbf{G}) P_{nlm}^{I}(\mathbf{G})\Bigg] h_{l}^{n,n';I} \Bigg[\sum_{\mathbf{G}'} P_{n'lm}^{I}(\mathbf{G}')\psi_q(\mathbf{G}')\Bigg]
\end{equation}
where $P_{nlm}^{I}(\mathbf{G})$ is the reciprocal space representation of the nonlocal projector obtained from the Kleinman-Bylander (or generalized Kleinman-Bylander~\cite{vanderbilt1985optimally,blochl1990generalized}) expansion of the pseudopotential, which can be obtained from spherical Bessel transforms.
\begin{equation*}
P_{nlm}^{I}(\mathbf{G}) = \frac{4\pi}{\Omega} e^{ \mathit{i} \mathbf{G} \cdot \mathbf{R}_I} \mathit{i}^{-l} T_{l,m}(\mathbf{G}) \int_0^{\infty} P_{nlm}^I(r) j_l(|\mathbf{G}|r) r^2 dr 
\end{equation*}
where $T_{l,m}$ is a real space spherical harmonic or Tesseral harmonic~\cite{bylaska2020filon}, and  $j_l(x)$ are the spherical Bessel functions of the first kind of degree $l$.

\subsection{Periodic Two-Electron Integrals using the Pseudopotential Plane-Wave Method}
\label{subsec:2-electron-pspw}
The two-electron integrals are written as
\begin{equation}
h_{pqrs} = \left\lbrace 
\begin{array}{ll} 
2 E_\text{periodic}^{pqrs} + 2 E_{\text{exch}}^{pqrs}  &  \text{if \textit{p=q} or \textit{r=s}} \\ 
2 E_\text{screened}^{pqrs}  & \text{otherwise} \end{array} \right.
\label{eqn:2e}
\end{equation}
where the periodic Coulomb and screened Coulomb energies are
\begin{equation*}
E_\text{periodic}^{pqrs} = \frac{\Omega}{2} \sum_{\mathbf{G\neq0}}  \rho^{*}_{pq}(\mathbf{G}) \frac{4\pi}{|\mathbf{G}|^2}  \rho_{rs}(\mathbf{G})
\end{equation*}
and
\begin{equation*}
E^{pqrs}_\text{screened} = \frac{\Omega}{2} \sum_{\mathbf{G}}
\rho^{*}_{pq}(\mathbf{G}) \rho_{rs}(\mathbf{G})
V_{f}(\mathbf{G})
\end{equation*}
where the Fourier representation of the densities are
\begin{equation*}
\rho_{pq}(\mathbf{G})=\sum_{\mathbf{G}'}\psi_{p}^*(\mathbf{G}')\psi_{q}(\mathbf{G}'+\mathbf{G}).
\label{rho_G}
\end{equation*}
The periodic exchange term in Eq.~\ref{eqn:2e} is approximated by
\begin{equation*}
E^{pqrs}_{\text{exch}} \approx - E^{pqrs}_\text{screened}.
\end{equation*}
The filter potential is approximated using the cutoff Coulomb kernel 
from our prior exchange paper based on the Wannier orbitals~\cite{bylaska2011parallel}, written in real-space as
\begin{equation*}
    V_{f}(R) = \frac{1-\left(1-e^{-\left(\frac{R}{R_\text{cut}}\right)^{N+2}}\right)^{N}}{R},
\end{equation*}
where $R = |{\bf r}_i - {\bf r}_j|$, and $N$ and $R_{cut}$ are adjustable parameters. 
The design of this cutoff kernel is chosen to remove the interactions between redundant periodic images of Wannier orbitals, because of the long-range nature of the Coulomb potential.  Recently, we developed a Filon integration strategy~\cite{bylaska2020filon}, which showed that filter potential for periodic exchange can be formulated as
\begin{equation*}
V_{f}(\mathbf{G}) = \frac{1}{(\Omega V_\text{BZ})^2} \iint_{\bar{V}_{BZ}}
\frac{4\pi}{|\mathbf{G}-\mathbf{k}+\mathbf{l}|^2} d\mathbf{k} d\mathbf{l},
\label{hf_kernel}
\end{equation*}
where $V_\text{BZ}=\frac{2\pi^2}{\Omega}$ is the volume of the first Brillouin zone, and moreover this potential can be approximated by the cutoff Coulomb kernel. 

To derive the form of the Eqs.~\ref{eqn:1e} and~\ref{eqn:2e}, we compared the results from the ”corresponding orbital transformation” developed by King et al.~\cite{king1967corresponding} (and generalized to periodic boundary conditions, see section~\ref{sec:periodic_covos} and reference ~\cite{bylaska2018corresponding}) to the results using the one-electron and two-electron integrals for the electronic structure Hamiltonian integrals, $H_{AB}= \braket{\Psi_A|H|\Psi_B}$ between two determinants $\ket{\Psi_A}$ and $\ket{\Psi_B}$.

\subsection{Periodic Ion-Ion Electrostatic Energy using the Pseudopotential Plane-Wave Method}
\label{subsection:ewald}
The ion-ion electrostatic energy for a periodic system can be calculated using the Ewald decomposition~\cite{ewald1921vii}.
\begin{eqnarray}
\label{eqn:ewald}
E_\text{electrostatic}^\text{ion-ion} &=& \frac{1}{2\Omega} \sum_{\mathbf{G} \neq 0} \frac{4\pi}{|\mathbf{G}|^2} \exp \left(- \frac{|\mathbf{G}|^2}{4\epsilon} \right) \nonumber \\
& \times & \left[ \sum_{I,J} Z_I \exp \left(\mathit{i} \mathbf{G} \cdot \mathbf{R}_I \right) Z_J \exp \left(-\mathit{i} \mathbf{G} \cdot \mathbf{R}_J \right) \right] \nonumber \\
&+& \frac{1}{2} \sum_{\mathbf{L}} \sum_{I,J \in |\mathbf{R}_I \mathbf{R}_J + \mathbf{L}| \neq 0} Z_I Z_J \frac{\mathrm{erfc} ( \epsilon |\mathbf{R}_I - \mathbf{R}_J + \mathbf{L}|) }{|\mathbf{R}_I - \mathbf{R}_J + \mathbf{L}|} \nonumber \\
& - & \frac{\epsilon}{\sqrt{\pi}} \sum_I Z_I^2 - \frac{\pi}{2\epsilon^2\Omega} \left( \sum_I Z_I \right)^2 
\end{eqnarray}
where $Z_I$ are atom charges, $\epsilon$ is a constant (typically on the order of 1) and $\mathbf{L}$ is a lattice vector.

\section{Algorithm for defining a virtual space with a small CI Hamiltonian}
\label{sec:periodic_covos}
In this section, we present a downfolding method to define virtual orbitals for expanding the second-quantized Hamiltonian given in Eq.~\ref{eqn:secondH}.  As previously shown, these new types of orbitals are able to capture significantly more correlation energy than with virtual orbitals coming from Hartree--Fock~\cite{bylaska2021quantum}.  The basis of this method is to define a set of virtual orbitals, $\{\psi^{(n)}_e(\mathbf{r})\}$ with $n=1..N_{virtual}$, which we call correlation optimized virtual orbitals or COVOs for short, by optimizing a small select configuration interaction (CI) Hamiltonian with respect to a single virtual orbital, and then the next virtual orbitals in sequence, subject to them being orthonormal to the filled and previously computed virtual orbitals.
The algorithm to calculate these new type of orbitals can be formulated as follows:
\begin{enumerate}
    \item Set $n=1$
    \item Using the ground state one-electron orbitals for many electron systems, $\psi_{f_1}(\mathbf{r})$, $\psi_{f_2}(\mathbf{r})$, $\cdots$, $\psi_{f_N}(\mathbf{r})$, and the virtual orbital to be optimized, $\psi^{(n)}_e(\mathbf{r})$, generate a CI matrix. 
    \item Calculate the select CI expansion coefficients by diagonalizing the CI matrix.
    \item Using the CI coefficients associated with the lowest eigenvalue, calculate the gradient with respect to the $\psi^{(n)}_e(\mathbf{r})$ then update with a conjugate gradient or similar method while making sure that $\psi^{(n)}_e(\mathbf{r})$ is normalized and orthogonal to $\psi_{f_1}(\mathbf{r})$, $\psi_{f_2}(\mathbf{r})$, $\cdots$, $\psi_{f_N}(\mathbf{r})$ and $\psi^{(m)}_e(\mathbf{r})$ for $m=1,...,n-1$.
    \item If the gradient is small then $n=n+1$
    \item If $n\leq N_{virtual}$ go to step 2, otherwise finished.
\end{enumerate}

A small CI wavefunction is constructed by varying the top orbitals to produce 3 determinant wavefunctions for the 2N-electron system composed of (N+1) one-electron orbitals, $\psi_{f_1}(\mathbf{r})$ and $\psi^{(n)}_e(\mathbf{r})$, can be written as a linear combination of 6 determinant wavefunctions, or just 3 determinant wavefunctions for just singlet (or triplet) states. 
\begin{eqnarray*}
\Psi_i[\psi_{f_1}(\mathbf{r}), \cdots, \psi_{f_{N-1}}(\mathbf{r}), \psi_{f_N}(\mathbf{r}), \psi_e(\mathbf{r})] 
&=& c_g^{(i)} \Psi_{g}[\psi_{f_1}(\mathbf{r}), \cdots, \psi_{f_{N-1}}(\mathbf{r}), \psi_{f_N}(\mathbf{r})] \\
&+& c_e^{(i)} \Psi_{e}[\psi_{f_1}(\mathbf{r}), \cdots, \psi_{f_{N-1}}(\mathbf{r}), \psi_{e}(\mathbf{r})]   \\
&+& c_m^{(i)} \Psi_{m}[\psi_{f_1}(\mathbf{r}), \cdots, \psi_{f_{N-1}}(\mathbf{r}), \psi_{f_N}(\mathbf{r}), \psi_e(\mathbf{r})] + \cdots  \\
\end{eqnarray*}
Using this small CI ansatz, the energies of the system can be obtained by diagonalizing the following eigenvalue equation.
\begin{equation*}
\label{3x3eig}
H C_i = E_i S C_i
\end{equation*}
where 
\begin{eqnarray}
H & = & 
\begin{bmatrix}
    \left\langle\Psi_g | H | \Psi_g\right\rangle   & 
    \left\langle\Psi_g | H | \Psi_e\right\rangle   &
    \left\langle\Psi_g | H | \Psi_m\right\rangle   \\
    \left\langle\Psi_e | H | \Psi_g\right\rangle    & 
    \left\langle\Psi_e | H | \Psi_e\right\rangle    &
    \left\langle\Psi_e | H | \Psi_m\right\rangle \\
    \left\langle\Psi_m | H | \Psi_g\right\rangle    & 
    \left\langle\Psi_m | H | \Psi_e\right\rangle    &
    \left\langle\Psi_m | H | \Psi_m\right\rangle \\
\end{bmatrix} \label{eqn:Hmatrix} \\
S & = &
\begin{bmatrix}
    \left\langle\Psi_g  | \Psi_g\right\rangle   & 
    \left\langle\Psi_g  | \Psi_e\right\rangle   &
    \left\langle\Psi_g  | \Psi_m\right\rangle   \\
    \left\langle\Psi_e  | \Psi_g\right\rangle    & 
    \left\langle\Psi_e  | \Psi_e\right\rangle    &
    \left\langle\Psi_e  | \Psi_m\right\rangle \\
    \left\langle\Psi_m  | \Psi_g\right\rangle    & 
    \left\langle\Psi_m  | \Psi_e\right\rangle    &
    \left\langle\Psi_m  | \Psi_m\right\rangle \\
\end{bmatrix} \nonumber \\
C_i & = &
\begin{bmatrix}
    c_{g}^{(i)} \\
    c_{e}^{(i)}  \\ 
    c_{m}^{(i)}
\end{bmatrix} \nonumber
\end{eqnarray}

Note the overlap matrix, $S$, is the identity matrix for orthonormal $\psi_g$ and $\psi_e$. The variation with respect to $\psi_e(\mathbf{r})$ can simply be obtained using the following formula.
\begin{eqnarray} \label{eqn:psiegradient}
\cfrac{\delta E_i}{\delta \psi^{*}_e(\mathbf{r})} 
&=& c_{g}^{(i)} \cfrac{\delta \left\langle\Psi_g |H|\Psi_g\right\rangle}{\delta \psi^{*}_e(\mathbf{r})} c_{g}^{(i)}  
+ c_{g}^{(i)} \cfrac{\delta \left\langle\Psi_g |H|\Psi_e\right\rangle}{\delta \psi^{*}_e(\mathbf{r})} c_{e}^{(i)} \nonumber \\  
&+& c_{g}^{(i)} \cfrac{\delta \left\langle\Psi_g |H|\Psi_m\right\rangle}{\delta \psi^{*}_e(\mathbf{r})} c_{m}^{(i)}   
+ c_{e}^{(i)} \cfrac{\delta \left\langle\Psi_e |H|\Psi_g\right\rangle}{\delta \psi^{*}_e(\mathbf{r})} c_{g}^{(i)} \nonumber \\  
&+& c_{e}^{(i)} \cfrac{\delta \left\langle\Psi_e |H|\Psi_e\right\rangle}{\delta \psi^{*}_e(\mathbf{r})} c_{e}^{(i)}   
+ c_{e}^{(i)} \cfrac{\delta \left\langle\Psi_e |H|\Psi_m\right\rangle}{\delta \psi^{*}_e(\mathbf{r})} c_{m}^{(i)} \nonumber \\
&+& c_{m}^{(i)} \cfrac{\delta \left\langle\Psi_m |H|\Psi_g\right\rangle}{\delta \psi^{*}_e(\mathbf{r})} c_{g}^{(i)}   
+ c_{m}^{(i)} \cfrac{\delta \left\langle\Psi_m |H|\Psi_e\right\rangle}{\delta \psi^{*}_e(\mathbf{r})} c_{e}^{(i)} \nonumber \\  
&+& c_{m}^{(i)} \cfrac{\delta \left\langle\Psi_m |H|\Psi_m\right\rangle}{\delta \psi^{*}_e(\mathbf{r})} c_{m}^{(i)}  
\end{eqnarray}

It should be noted that the above formulas can be generalized to work beyond two electron systems by using corresponding orbitals techniques~\cite{king1967corresponding,bylaska2018corresponding}.  The next two subsections, \ref{sec:one-electron}-\ref{sec:two-electron-matrix}, provide formulas that can be used to generate the matrix elements in Eq.~\ref{eqn:Hmatrix} and the gradients with respect to $\psi^{*}_e(\mathbf{r})$ in Eq.~\ref{eqn:psiegradient}.

We also note the COVOs approach is similar in spirit to the optimized virtual orbital space (OVOS) approach developed over 30 years ago by Adamowicz and Bartlett~\cite{adamowicz1987optimized,adamowicz1988optimized}. The main difference is that the variational space used by COVOs is significantly larger, because plane-wave basis sets are used instead of LCAO Gaussian basis sets used by OVOS.  Another difference between the approaches is that the correlation is described by a small CI Hamiltonian for COVOs, and a
second-order Hylleraas functional~\cite{hylleraas1928grundzustand,hylleraas1929neue,hylleraas1930grundterm,hylleraas1964schrodinger,koga1992hylleraas} for OVOS.  The cost to generate COVOs is similar to the cost to generate regular RHF virtual orbitals (just 4 to 9 times more expensive than RHF virtual orbitals). However, because the orbitals are generated one at a time, the resulting electronic gradient is non-Hermitian, which requires more advanced optimizers.

\vspace{1cm}
\subsection{One-electron Virtual and N-Filled Orbitals}
\label{sec:one-electron}

The one-electron spin orbitals of (N+1)-state Hamiltonian are
\begin{center}
\begin{tikzpicture}
  \draw[loosely dotted, thick] (-1.5,-4.1)--(-1.5,-3.5);
  \drawLevel[elec = up, pos = {(-2.0,-2.8)}, width = 1.0,spinstyle={very thick,color=blue,->}]{label1}
  \draw[loosely dotted, thick] (-1.5,-2.0)--(-1.5,-1.4);
  \drawLevel[elec = no, pos = {(-2.0,-0.6)}, width = 1.0,spinstyle={very thick,color=red,->}]{label2}
  \drawLevel[elec = no, pos = {(-2.0,0.6)}, width = 1.0,spinstyle={very thick,color=red,->}]{label3}
  
  \draw[loosely dotted, thick] (0.5,-4.1)--(0.5,-3.5);
  \drawLevel[elec = down, pos = {(0.0,-2.8)}, width = 1.0,spinstyle={very thick,color=blue,->}]{}
  \draw[loosely dotted, thick] (0.5,-2.0)--(0.5,-1.4);
  \drawLevel[elec = no, pos = {(0.0,-0.6)}, width = 1.0,spinstyle={very thick,color=red,->}]{}
  \drawLevel[elec = no, pos = {(0.0,0.6)}, width = 1.0,spinstyle={very thick,color=red,->}]{}
  
  \draw[loosely dotted, thick] (2.5,-4.1)--(2.5,-3.5);
  \drawLevel[elec = no, pos = {(2,-2.8)}, width = 1.0,spinstyle={very thick,color=blue,->}]{}
  \draw[loosely dotted, thick] (2.5,-2.0)--(2.5,-1.4);
  \drawLevel[elec = up, pos = {(2,-0.6)}, width = 1,spinstyle={very thick,color=red,->}]{}
  \drawLevel[elec = no, pos = {(2,0.6)}, width = 1,spinstyle={very thick,color=red,->}]{}
  
  \draw[loosely dotted, thick] (4.5,-4.1)--(4.5,-3.5);
  \drawLevel[elec = no, pos = {(4,-2.8)}, width = 1.0,spinstyle={very thick,color=blue,->}]{}
  \draw[loosely dotted, thick] (4.5,-2.0)--(4.5,-1.4);
  \drawLevel[elec = down, pos = {(4,-0.6)}, width = 1,spinstyle={very thick,color=red,->}]{}
  \drawLevel[elec = no, pos = {(4,0.6)}, width = 1,spinstyle={very thick,color=red,->}]{}
  
  \draw[loosely dotted, thick] (6.5,-4.1)--(6.5,-3.5);
  \drawLevel[elec = no, pos = {(6,-2.8)}, width = 1.0,spinstyle={very thick,color=blue,->}]{}
  \draw[loosely dotted, thick] (6.5,-2.0)--(6.5,-1.4);
  \drawLevel[elec = no, pos = {(6.0,-0.6)}, width = 1,spinstyle={very thick,color=red,->}]{}
  \drawLevel[elec = up, pos = {(6.0,0.6)}, width = 1,spinstyle={very thick,color=red,->}]{}
  
  \draw[loosely dotted, thick] (8.5,-4.1)--(8.5,-3.5);
  \drawLevel[elec = no, pos = {(8,-2.8)}, width = 1.0,spinstyle={very thick,color=blue,->}]{}
  \draw[loosely dotted, thick] (8.5,-2.0)--(8.5,-1.4);
  \drawLevel[elec = no, pos = {(8.0,-0.6)}, width = 1,spinstyle={very thick,color=red,->}]{}
  \drawLevel[elec = down, pos = {(8.0,0.6)}, width = 1,spinstyle={very thick,color=red,->}]{}

  \node[left] at (left label1) {$f_i$};
  \node[left] at (left label2) {$f_N$};
  \node[left] at (left label3) {$e$};
  \node[above] at (-1.5,1.25) {$\ket{\chi_{f_i}^\alpha}$};
  \node[above] at (0.5,1.25) {$\ket{\chi_{f_i}^\beta}$};
  \node[above] at (2.5,1.25) {$\ket{\chi_{f_N}^\alpha}$};
  \node[above] at (4.5,1.25) {$\ket{\chi_{f_N}^\beta}$};
  \node[above] at (6.5,1.25) {$\ket{\chi_e^\alpha}$};
  \node[above] at (8.5,1.25) {$\ket{\chi_e^\beta}$};
\end{tikzpicture}
\end{center}
\begin{eqnarray*}
    \chi_{f_i}^\alpha(\mathbf{x}) &=& \psi_{f_i}(\mathbf{r}) \alpha(s), \: \forall_{i=1,N}\\
    \chi_{f_i}^\beta(\mathbf{x}) &=& \psi_{f_i}(\mathbf{r}) \beta(s), \: \forall_{i=1,N}\\
    \chi_e^\alpha(\mathbf{x}) &=& \psi_e(\mathbf{r}) \alpha(s)\\
    \chi_e^\beta(\mathbf{x}) &=& \psi_e(\mathbf{r}) \beta(s)
\end{eqnarray*}
where the spatial orbitals and spin functions are orthonormalized.

\begin{equation*}
  \int \psi^{*}_{f_i}(\mathbf{r})  \psi_e(\mathbf{r}) d\mathbf{r} = 
  \int \psi^{*}_e(\mathbf{r})  \psi_{f_i}(\mathbf{r}) d\mathbf{r} = 0
\end{equation*}

\begin{equation*}
  \int \psi^{*}_e(\mathbf{r})  \psi_e(\mathbf{r}) d\mathbf{r} = 1
\end{equation*}

\begin{equation*}
  \int \psi^{*}_{f_i}(\mathbf{r})  \psi_{f_j}(\mathbf{r}) d\mathbf{r} = \delta_{ij}
\end{equation*}

\begin{equation*}
   \int \alpha^{*}(s)  \beta(s) ds = \int \beta^{*}(s) \alpha(s) ds =  0  
\end{equation*}

\begin{equation*}
   \int \alpha^{*}(s)  \alpha(s) ds =  \int \beta^{*}(s)  \beta(s) ds = 1  
\end{equation*}

\vspace{1cm}
\subsection{The 2N-electron Determinants of an (N+1)-state Hamiltonian With Different Top Level Fillings}
\label{sec:two-electron}
For the (N+1)-state system, there are six 2N-electron wavefunctions, two of which are singlet, two of which are triplet, and two of which contain a mixture of singlet and triplet character.  These wavefunctions can be written as
\newline
\begin{center}
\begin{tikzpicture}
  \drawLevel[elec = pair, pos = {(-2,-2.4)}, width = 1, spinstyle={very thick,color=blue,->}]{label1}
  \draw[loosely dotted, thick] (-1.5,-1.8)--(-1.5,-1.2);
  \drawLevel[elec = pair, pos = {(-2,-0.6)}, width = 1, spinstyle={very thick,color=blue,->}]{label2}
  \drawLevel[elec = pair, pos = {(-2,0.6)}, width = 1, spinstyle={very thick,color=red,->}]{label3}
  \drawLevel[elec = no, pos = {(-2,1.8)}, width = 1, spinstyle={very thick,color=red,->}]{label4}  
  
  \drawLevel[elec = pair, pos = {(0,-2.4)}, width = 1, spinstyle={very thick,color=blue,->}]{}
  \draw[loosely dotted, thick] (0.5,-1.8)--(0.5,-1.2);
  \drawLevel[elec = pair, pos = {(0,-0.6)}, width = 1, spinstyle={very thick,color=blue,->}]{}
  \drawLevel[elec = no, pos = {(0,0.6)}, width = 1, spinstyle={very thick,color=red,->}]{}
  \drawLevel[elec = pair, pos = {(0,1.8)}, width = 1, spinstyle={very thick,color=red,->}]{}
  
  \drawLevel[elec = pair, pos = {(2,-2.4)}, width = 1, spinstyle={very thick,color=blue,->}]{}
  \draw[loosely dotted, thick] (2.5,-1.8)--(2.5,-1.2);
  \drawLevel[elec = pair, pos = {(2,-0.6)}, width = 1, spinstyle={very thick,color=blue,->}]{}
  \drawLevel[elec = up, pos = {(2,0.6)}, width = 1, spinstyle={very thick,color=red,->}]{}
  \drawLevel[elec = down, pos = {(2,1.8)}, width = 1, spinstyle={very thick,color=red,->}]{}
  
  \drawLevel[elec = pair, pos = {(4,-2.4)}, width = 1, spinstyle={very thick,color=blue,->}]{}
  \draw[loosely dotted, thick] (4.5,-1.8)--(4.5,-1.2);
  \drawLevel[elec = pair, pos = {(4,-0.6)}, width = 1, spinstyle={very thick,color=blue,->}]{}
  \drawLevel[elec = down, pos = {(4,0.6)}, width = 1, spinstyle={very thick,color=red,->}]{}
  \drawLevel[elec = up, pos = {(4,1.8)}, width = 1, spinstyle={very thick,color=red,->}]{}

  \drawLevel[elec = pair, pos = {(6,-2.4)}, width = 1, spinstyle={very thick,color=blue,->}]{}
  \draw[loosely dotted, thick] (6.5,-1.8)--(6.5,-1.2);
  \drawLevel[elec = pair, pos = {(6,-0.6)}, width = 1, spinstyle={very thick,color=blue,->}]{}
  \drawLevel[elec = up, pos = {(6,0.6)}, width = 1, spinstyle={very thick,color=red,->}]{}
  \drawLevel[elec = up, pos = {(6,1.8)}, width = 1, spinstyle={very thick,color=red,->}]{}

  \drawLevel[elec = pair, pos = {(8,-2.4)}, width = 1, spinstyle={very thick,color=blue,->}]{}
  \draw[loosely dotted, thick] (8.5,-1.8)--(8.5,-1.2);
  \drawLevel[elec = pair, pos = {(8,-0.6)}, width = 1, spinstyle={very thick,color=blue,->}]{}
  \drawLevel[elec = down, pos = {(8,0.6)}, width = 1, spinstyle={very thick,color=red,->}]{}
  \drawLevel[elec = down, pos = {(8,1.8)}, width = 1, spinstyle={very thick,color=red,->}]{}

  \node[left] at (left label1) {$f_1$};
  \node[left] at (left label2) {$f_{N-1}$};
  \node[left] at (left label3) {$f_N$};
  \node[left] at (left label4) {$e$};
  \node[above] at (-1.5,2.5) {$\ket{\Psi_g}$};
  \node[above] at (0.5,2.5) {$\ket{\Psi_e}$};
  \node[above] at (2.5,2.5) {$\ket{\Psi_a}$};
  \node[above] at (4.5,2.5) {$\ket{\Psi_b}$};
  \node[above] at (6.5,2.5) {$\ket{\Psi_u}$};
  \node[above] at (8.5,2.5) {$\ket{\Psi_d}$};
\end{tikzpicture}
\end{center}

\begin{eqnarray*}
\Psi_g(\mathbf{x}_1,\mathbf{x}_2, \cdots, \mathbf{x}_{2N-1},\mathbf{x}_{2N}) &=& \frac{1}{\sqrt{(2N)!}}
\begin{vmatrix}
\psi_{f_1}(\mathbf{r}_1)\alpha(s_1) & \psi_{f_1}(\mathbf{r}_2)\alpha(s_2) & \cdots & 
\psi_{f_1}(\mathbf{r}_{2N})\alpha(s_{2N})\\
\psi_{f_1}(\mathbf{r}_1)\beta(s_1) & \psi_{f_1}(\mathbf{r}_2)\beta(s_2) & \cdots & 
\psi_{f_1}(\mathbf{r}_{2N})\beta(s_{2N})\\
\vdots & \vdots & \vdots & \vdots\\
\psi_{f_{N-1}}(\mathbf{r}_1)\alpha(s_1) & \psi_{f_{N-1}}(\mathbf{r}_2)\alpha(s_2) & \cdots & 
\psi_{f_{N-1}}(\mathbf{r}_{2N})\alpha(s_{2N})\\
\psi_{f_{N-1}}(\mathbf{r}_1)\beta(s_1) & \psi_{f_{N-1}}(\mathbf{r}_2)\beta(s_2) & \cdots & 
\psi_{f_{N-1}}(\mathbf{r}_{2N})\beta(s_{2N})\\
\psi_{f_N}(\mathbf{r}_1)\alpha(s_1) & \psi_{f_N}(\mathbf{r}_2)\alpha(s_2) & \cdots & 
\psi_{f_N}(\mathbf{r}_{2N})\alpha(s_{2N})\\
\psi_{f_N}(\mathbf{r}_1)\beta(s_1) & \psi_{f_N}(\mathbf{r}_2)\beta(s_2) & \cdots & 
\psi_{f_N}(\mathbf{r}_{2N})\beta(s_{2N})
\end{vmatrix}\\
&\estimates& \ket{\chi_{f_1}^\alpha \chi_{f_1}^\beta \cdots \chi_{f_{N-1}}^\alpha \chi_{f_{N-1}}^\beta \chi_{f_N}^\alpha \chi_{f_N}^\beta} 
\end{eqnarray*}

\begin{eqnarray*}
\Psi_e(\mathbf{x}_1,\mathbf{x}_2, \cdots, \mathbf{x}_{2N-1},\mathbf{x}_{2N}) &=& \frac{1}{\sqrt{(2N)!}}
\begin{vmatrix}
\psi_{f_1}(\mathbf{r}_1)\alpha(s_1) & \psi_{f_1}(\mathbf{r}_2)\alpha(s_2) & \cdots & 
\psi_{f_1}(\mathbf{r}_{2N})\alpha(s_{2N})\\
\psi_{f_1}(\mathbf{r}_1)\beta(s_1) & \psi_{f_1}(\mathbf{r}_2)\beta(s_2) & \cdots & 
\psi_{f_1}(\mathbf{r}_{2N})\beta(s_{2N})\\
\vdots & \vdots & \vdots & \vdots\\
\psi_{f_{N-1}}(\mathbf{r}_1)\alpha(s_1) & \psi_{f_{N-1}}(\mathbf{r}_2)\alpha(s_2) & \cdots & 
\psi_{f_{N-1}}(\mathbf{r}_{2N})\alpha(s_{2N})\\
\psi_{f_{N-1}}(\mathbf{r}_1)\beta(s_1) & \psi_{f_{N-1}}(\mathbf{r}_2)\beta(s_2) & \cdots & 
\psi_{f_{N-1}}(\mathbf{r}_{2N})\beta(s_{2N})\\
\psi_{e}(\mathbf{r}_1)\alpha(s_1) & \psi_{e}(\mathbf{r}_2)\alpha(s_2) & \cdots & 
\psi_{e}(\mathbf{r}_{2N})\alpha(s_{2N})\\
\psi_{e}(\mathbf{r}_1)\beta(s_1) & \psi_{e}(\mathbf{r}_2)\beta(s_2) & \cdots & 
\psi_{e}(\mathbf{r}_{2N})\beta(s_{2N})
\end{vmatrix}\\
&\estimates& \ket{\chi_{f_1}^\alpha \chi_{f_1}^\beta \cdots \chi_{f_{N-1}}^\alpha \chi_{f_{N-1}}^\beta \chi_e^\alpha \chi_e^\beta} 
\end{eqnarray*}

\begin{eqnarray*}
\Psi_a(\mathbf{x}_1,\mathbf{x}_2, \cdots, \mathbf{x}_{2N-1},\mathbf{x}_{2N}) &=& \frac{1}{\sqrt{(2N)!}}
\begin{vmatrix}
\psi_{f_1}(\mathbf{r}_1)\alpha(s_1) & \psi_{f_1}(\mathbf{r}_2)\alpha(s_2) & \cdots & 
\psi_{f_1}(\mathbf{r}_{2N})\alpha(s_{2N})\\
\psi_{f_1}(\mathbf{r}_1)\beta(s_1) & \psi_{f_1}(\mathbf{r}_2)\beta(s_2) & \cdots & 
\psi_{f_1}(\mathbf{r}_{2N})\beta(s_{2N})\\
\vdots & \vdots & \vdots & \vdots\\
\psi_{f_{N-1}}(\mathbf{r}_1)\alpha(s_1) & \psi_{f_{N-1}}(\mathbf{r}_2)\alpha(s_2) & \cdots & 
\psi_{f_{N-1}}(\mathbf{r}_{2N})\alpha(s_{2N})\\
\psi_{f_{N-1}}(\mathbf{r}_1)\beta(s_1) & \psi_{f_{N-1}}(\mathbf{r}_2)\beta(s_2) & \cdots & 
\psi_{f_{N-1}}(\mathbf{r}_{2N})\beta(s_{2N})\\
\psi_{f_N}(\mathbf{r}_1)\alpha(s_1) & \psi_{f_N}(\mathbf{r}_2)\alpha(s_2) & \cdots & 
\psi_{f_N}(\mathbf{r}_{2N})\alpha(s_{2N})\\
\psi_{e}(\mathbf{r}_1)\beta(s_1) & \psi_{e}(\mathbf{r}_2)\beta(s_2) & \cdots & 
\psi_{e}(\mathbf{r}_{2N})\beta(s_{2N})
\end{vmatrix}\\
&\estimates& \ket{\chi_{f_1}^\alpha \chi_{f_1}^\beta \cdots \chi_{f_{N-1}}^\alpha \chi_{f_{N-1}}^\beta \chi_{f_N}^\alpha \chi_e^\beta}
\end{eqnarray*}

\begin{eqnarray*}
\Psi_b(\mathbf{x}_1,\mathbf{x}_2, \cdots, \mathbf{x}_{2N-1},\mathbf{x}_{2N}) &=& \frac{1}{\sqrt{(2N)!}}
\begin{vmatrix}
\psi_{f_1}(\mathbf{r}_1)\alpha(s_1) & \psi_{f_1}(\mathbf{r}_2)\alpha(s_2) & \cdots & 
\psi_{f_1}(\mathbf{r}_{2N})\alpha(s_{2N})\\
\psi_{f_1}(\mathbf{r}_1)\beta(s_1) & \psi_{f_1}(\mathbf{r}_2)\beta(s_2) & \cdots & 
\psi_{f_1}(\mathbf{r}_{2N})\beta(s_{2N})\\
\vdots & \vdots & \vdots & \vdots\\
\psi_{f_{N-1}}(\mathbf{r}_1)\alpha(s_1) & \psi_{f_{N-1}}(\mathbf{r}_2)\alpha(s_2) & \cdots & 
\psi_{f_{N-1}}(\mathbf{r}_{2N})\alpha(s_{2N})\\
\psi_{f_{N-1}}(\mathbf{r}_1)\beta(s_1) & \psi_{f_{N-1}}(\mathbf{r}_2)\beta(s_2) & \cdots & 
\psi_{f_{N-1}}(\mathbf{r}_{2N})\beta(s_{2N})\\
\psi_{f_N}(\mathbf{r}_1)\beta(s_1) & \psi_{f_N}(\mathbf{r}_2)\beta(s_2) & \cdots & 
\psi_{f_N}(\mathbf{r}_{2N})\beta(s_{2N})\\
\psi_{e}(\mathbf{r}_1)\alpha(s_1) & \psi_{e}(\mathbf{r}_2)\alpha(s_2) & \cdots & 
\psi_{e}(\mathbf{r}_{2N})\alpha(s_{2N})
\end{vmatrix}\\
&\estimates& \ket{\chi_{f_1}^\alpha \chi_{f_1}^\beta \cdots \chi_{f_{N-1}}^\alpha \chi_{f_{N-1}}^\beta \chi_{f_N}^\beta \chi_e^\alpha}
\end{eqnarray*}

\begin{eqnarray*}
\Psi_u(\mathbf{x}_1,\mathbf{x}_2, \cdots, \mathbf{x}_{2N-1},\mathbf{x}_{2N}) &=& \frac{1}{\sqrt{(2N)!}}
\begin{vmatrix}
\psi_{f_1}(\mathbf{r}_1)\alpha(s_1) & \psi_{f_1}(\mathbf{r}_2)\alpha(s_2) & \cdots & 
\psi_{f_1}(\mathbf{r}_{2N})\alpha(s_{2N})\\
\psi_{f_1}(\mathbf{r}_1)\beta(s_1) & \psi_{f_1}(\mathbf{r}_2)\beta(s_2) & \cdots & 
\psi_{f_1}(\mathbf{r}_{2N})\beta(s_{2N})\\
\vdots & \vdots & \vdots & \vdots\\
\psi_{f_{N-1}}(\mathbf{r}_1)\alpha(s_1) & \psi_{f_{N-1}}(\mathbf{r}_2)\alpha(s_2) & \cdots & 
\psi_{f_{N-1}}(\mathbf{r}_{2N})\alpha(s_{2N})\\
\psi_{f_{N-1}}(\mathbf{r}_1)\beta(s_1) & \psi_{f_{N-1}}(\mathbf{r}_2)\beta(s_2) & \cdots & 
\psi_{f_{N-1}}(\mathbf{r}_{2N})\beta(s_{2N})\\
\psi_{f_N}(\mathbf{r}_1)\alpha(s_1) & \psi_{f_N}(\mathbf{r}_2)\alpha(s_2) & \cdots & 
\psi_{f_N}(\mathbf{r}_{2N})\alpha(s_{2N})\\
\psi_{e}(\mathbf{r}_1)\alpha(s_1) & \psi_{e}(\mathbf{r}_2)\alpha(s_2) & \cdots & 
\psi_{e}(\mathbf{r}_{2N})\alpha(s_{2N})
\end{vmatrix}\\
&\estimates& \ket{\chi_{f_1}^\alpha \chi_{f_1}^\beta \cdots \chi_{f_{N-1}}^\alpha \chi_{f_{N-1}}^\beta \chi_{f_N}^\alpha \chi_e^\alpha}
\end{eqnarray*}

\begin{eqnarray*}
\Psi_d(\mathbf{x}_1,\mathbf{x}_2, \cdots, \mathbf{x}_{2N-1},\mathbf{x}_{2N}) &=& \frac{1}{\sqrt{(2N)!}}
\begin{vmatrix}
\psi_{f_1}(\mathbf{r}_1)\alpha(s_1) & \psi_{f_1}(\mathbf{r}_2)\alpha(s_2) & \cdots & 
\psi_{f_1}(\mathbf{r}_{2N})\alpha(s_{2N})\\
\psi_{f_1}(\mathbf{r}_1)\beta(s_1) & \psi_{f_1}(\mathbf{r}_2)\beta(s_2) & \cdots & 
\psi_{f_1}(\mathbf{r}_{2N})\beta(s_{2N})\\
\vdots & \vdots & \vdots & \vdots\\
\psi_{f_{N-1}}(\mathbf{r}_1)\alpha(s_1) & \psi_{f_{N-1}}(\mathbf{r}_2)\alpha(s_2) & \cdots & 
\psi_{f_{N-1}}(\mathbf{r}_{2N})\alpha(s_{2N})\\
\psi_{f_{N-1}}(\mathbf{r}_1)\beta(s_1) & \psi_{f_{N-1}}(\mathbf{r}_2)\beta(s_2) & \cdots & 
\psi_{f_{N-1}}(\mathbf{r}_{2N})\beta(s_{2N})\\
\psi_{f_N}(\mathbf{r}_1)\beta(s_1) & \psi_{f_N}(\mathbf{r}_2)\beta(s_2) & \cdots & 
\psi_{f_N}(\mathbf{r}_{2N})\beta(s_{2N})\\
\psi_{e}(\mathbf{r}_1)\beta(s_1) & \psi_{e}(\mathbf{r}_2)\beta(s_2) & \cdots & 
\psi_{e}(\mathbf{r}_{2N})\beta(s_{2N})
\end{vmatrix}\\
&\estimates& \ket{\chi_{f_1}^\alpha \chi_{f_1}^\beta \cdots \chi_{f_{N-1}}^\alpha \chi_{f_{N-1}}^\beta \chi_{f_N}^\beta \chi_e^\beta}
\end{eqnarray*}

Note that $\Psi_a$ and $\Psi_b$ cannot be written as a product of a spatial wavefunction times a spin-function.  Moreover, these functions are not eigenfunctions of the spin operators $S^2$ and $S_z$, and as a result these determinants contain both singlet and triplet components.  However, if we take linear combinations of them we can get two new wavefunctions that are separable in spatial and spin functions, and at the same time being eigenfunctions of $S^2$ and $S_z$.

\begin{eqnarray*}
\Psi_m (\mathbf{x}_1,\mathbf{x}_2, \cdots, \mathbf{x}_{2N-1},\mathbf{x}_{2N}) &=& \Psi_{a-b} (\mathbf{x}_1,\mathbf{x}_2, \cdots, \mathbf{x}_{2N-1},\mathbf{x}_{2N})\\
&=& \frac{1}{\sqrt{2}} \bigg[\Psi_{a} (\mathbf{x}_1,\mathbf{x}_2, \cdots, \mathbf{x}_{2N-1},\mathbf{x}_{2N}) - \Psi_{b} (\mathbf{x}_1,\mathbf{x}_2, \cdots, \mathbf{x}_{2N-1},\mathbf{x}_{2N}) \bigg]\\ 
&=& \frac{1}{\sqrt{2(2N)!}}
\begin{vmatrix}
\psi_{f_1}(\mathbf{r}_1)\alpha(s_1) & \psi_{f_1}(\mathbf{r}_2)\alpha(s_2) & \cdots & 
\psi_{f_1}(\mathbf{r}_{2N})\alpha(s_{2N})\\
\psi_{f_1}(\mathbf{r}_1)\beta(s_1) & \psi_{f_1}(\mathbf{r}_2)\beta(s_2) & \cdots & 
\psi_{f_1}(\mathbf{r}_{2N})\beta(s_{2N})\\
\vdots & \vdots & \vdots & \vdots\\
\psi_{f_{N-1}}(\mathbf{r}_1)\alpha(s_1) & \psi_{f_{N-1}}(\mathbf{r}_2)\alpha(s_2) & \cdots & 
\psi_{f_{N-1}}(\mathbf{r}_{2N})\alpha(s_{2N})\\
\psi_{f_{N-1}}(\mathbf{r}_1)\beta(s_1) & \psi_{f_{N-1}}(\mathbf{r}_2)\beta(s_2) & \cdots & 
\psi_{f_{N-1}}(\mathbf{r}_{2N})\beta(s_{2N})\\
\psi_{f_N}(\mathbf{r}_1)\alpha(s_1) & \psi_{f_N}(\mathbf{r}_2)\alpha(s_2) & \cdots & 
\psi_{f_N}(\mathbf{r}_{2N})\alpha(s_{2N})\\
\psi_{e}(\mathbf{r}_1)\beta(s_1) & \psi_{e}(\mathbf{r}_2)\beta(s_2) & \cdots & 
\psi_{e}(\mathbf{r}_{2N})\beta(s_{2N})
\end{vmatrix}\\
&-& \frac{1}{\sqrt{2(2N)!}}
\begin{vmatrix}
\psi_{f_1}(\mathbf{r}_1)\alpha(s_1) & \psi_{f_1}(\mathbf{r}_2)\alpha(s_2) & \cdots & 
\psi_{f_1}(\mathbf{r}_{2N})\alpha(s_{2N})\\
\psi_{f_1}(\mathbf{r}_1)\beta(s_1) & \psi_{f_1}(\mathbf{r}_2)\beta(s_2) & \cdots & 
\psi_{f_1}(\mathbf{r}_{2N})\beta(s_{2N})\\
\vdots & \vdots & \vdots & \vdots\\
\psi_{f_{N-1}}(\mathbf{r}_1)\alpha(s_1) & \psi_{f_{N-1}}(\mathbf{r}_2)\alpha(s_2) & \cdots & 
\psi_{f_{N-1}}(\mathbf{r}_{2N})\alpha(s_{2N})\\
\psi_{f_{N-1}}(\mathbf{r}_1)\beta(s_1) & \psi_{f_{N-1}}(\mathbf{r}_2)\beta(s_2) & \cdots & 
\psi_{f_{N-1}}(\mathbf{r}_{2N})\beta(s_{2N})\\
\psi_{f_N}(\mathbf{r}_1)\beta(s_1) & \psi_{f_N}(\mathbf{r}_2)\beta(s_2) & \cdots & 
\psi_{f_N}(\mathbf{r}_{2N})\beta(s_{2N})\\
\psi_{e}(\mathbf{r}_1)\alpha(s_1) & \psi_{e}(\mathbf{r}_2)\alpha(s_2) & \cdots & 
\psi_{e}(\mathbf{r}_{2N})\alpha(s_{2N})
\end{vmatrix}\\
&\estimates& \frac{1}{\sqrt{2}}\left(\ket{\chi_{f_1}^\alpha \chi_{f_1}^\beta \cdots \chi_{f_{N-1}}^\alpha \chi_{f_{N-1}}^\beta \chi_{f_N}^\alpha \chi_e^\beta} - \ket{\chi_{f_1}^\alpha \chi_{f_1}^\beta \cdots \chi_{f_{N-1}}^\alpha \chi_{f_{N-1}}^\beta \chi_{f_N}^\beta \chi_e^\alpha} \right)
\end{eqnarray*}

\begin{eqnarray*}
\Psi_p (\mathbf{x}_1,\mathbf{x}_2, \cdots, \mathbf{x}_{2N-1},\mathbf{x}_{2N}) &=& \Psi_{a+b} (\mathbf{x}_1,\mathbf{x}_2, \cdots, \mathbf{x}_{2N-1},\mathbf{x}_{2N})\\
&=& \frac{1}{\sqrt{2}} \bigg[\Psi_{a} (\mathbf{x}_1,\mathbf{x}_2, \cdots, \mathbf{x}_{2N-1},\mathbf{x}_{2N}) + \Psi_{b} (\mathbf{x}_1,\mathbf{x}_2, \cdots, \mathbf{x}_{2N-1},\mathbf{x}_{2N}) \bigg]\\ 
&=& \frac{1}{\sqrt{2(2N)!}}
\begin{vmatrix}
\psi_{f_1}(\mathbf{r}_1)\alpha(s_1) & \psi_{f_1}(\mathbf{r}_2)\alpha(s_2) & \cdots & 
\psi_{f_1}(\mathbf{r}_{2N})\alpha(s_{2N})\\
\psi_{f_1}(\mathbf{r}_1)\beta(s_1) & \psi_{f_1}(\mathbf{r}_2)\beta(s_2) & \cdots & 
\psi_{f_1}(\mathbf{r}_{2N})\beta(s_{2N})\\
\vdots & \vdots & \vdots & \vdots\\
\psi_{f_{N-1}}(\mathbf{r}_1)\alpha(s_1) & \psi_{f_{N-1}}(\mathbf{r}_2)\alpha(s_2) & \cdots & 
\psi_{f_{N-1}}(\mathbf{r}_{2N})\alpha(s_{2N})\\
\psi_{f_{N-1}}(\mathbf{r}_1)\beta(s_1) & \psi_{f_{N-1}}(\mathbf{r}_2)\beta(s_2) & \cdots & 
\psi_{f_{N-1}}(\mathbf{r}_{2N})\beta(s_{2N})\\
\psi_{f_N}(\mathbf{r}_1)\alpha(s_1) & \psi_{f_N}(\mathbf{r}_2)\alpha(s_2) & \cdots & 
\psi_{f_N}(\mathbf{r}_{2N})\alpha(s_{2N})\\
\psi_{e}(\mathbf{r}_1)\beta(s_1) & \psi_{e}(\mathbf{r}_2)\beta(s_2) & \cdots & 
\psi_{e}(\mathbf{r}_{2N})\beta(s_{2N})
\end{vmatrix}\\
&+& \frac{1}{\sqrt{2(2N)!}}
\begin{vmatrix}
\psi_{f_1}(\mathbf{r}_1)\alpha(s_1) & \psi_{f_1}(\mathbf{r}_2)\alpha(s_2) & \cdots & 
\psi_{f_1}(\mathbf{r}_{2N})\alpha(s_{2N})\\
\psi_{f_1}(\mathbf{r}_1)\beta(s_1) & \psi_{f_1}(\mathbf{r}_2)\beta(s_2) & \cdots & 
\psi_{f_1}(\mathbf{r}_{2N})\beta(s_{2N})\\
\vdots & \vdots & \vdots & \vdots\\
\psi_{f_{N-1}}(\mathbf{r}_1)\alpha(s_1) & \psi_{f_{N-1}}(\mathbf{r}_2)\alpha(s_2) & \cdots & 
\psi_{f_{N-1}}(\mathbf{r}_{2N})\alpha(s_{2N})\\
\psi_{f_{N-1}}(\mathbf{r}_1)\beta(s_1) & \psi_{f_{N-1}}(\mathbf{r}_2)\beta(s_2) & \cdots & 
\psi_{f_{N-1}}(\mathbf{r}_{2N})\beta(s_{2N})\\
\psi_{f_N}(\mathbf{r}_1)\beta(s_1) & \psi_{f_N}(\mathbf{r}_2)\beta(s_2) & \cdots & 
\psi_{f_N}(\mathbf{r}_{2N})\beta(s_{2N})\\
\psi_{e}(\mathbf{r}_1)\alpha(s_1) & \psi_{e}(\mathbf{r}_2)\alpha(s_2) & \cdots & 
\psi_{e}(\mathbf{r}_{2N})\alpha(s_{2N})
\end{vmatrix}\\
&\estimates& \frac{1}{\sqrt{2}}\left(\ket{\chi_{f_1}^\alpha \chi_{f_1}^\beta \cdots \chi_{f_{N-1}}^\alpha \chi_{f_{N-1}}^\beta \chi_{f_N}^\alpha \chi_e^\beta} + \ket{\chi_{f_1}^\alpha \chi_{f_1}^\beta \cdots \chi_{f_{N-1}}^\alpha \chi_{f_{N-1}}^\beta \chi_{f_N}^\beta \chi_e^\alpha} \right)
\end{eqnarray*}

\subsection{Incorporating Brillouin Zone Integration}
\label{subsec:transforming-pspw}

For systems with periodic boundary conditions, the matrix elements for calculating $H_{AB}$ are used with the Bloch states, i.e.
\begin{eqnarray*}
\mathbf{a}(\mathbf{x}) &= [ & a_{1\mathbf{k}_1}(\mathbf{x}),a_{2\mathbf{k}_1}(\mathbf{x}),\ldots,a_{N\mathbf{k}_1}(\mathbf{x}), \nonumber \\
& & a_{1\mathbf{k}_2}(\mathbf{x}),a_{2\mathbf{k}_2}(\mathbf{x}),\ldots,a_{N\mathbf{k}_2}(\mathbf{x}), \nonumber \\
& & \cdots ]^\dag \label{eqn: orb_a}\\
\mathbf{b}(\mathbf{x}) &= [ & b_{1\mathbf{k}_1}(\mathbf{x}),b_{2\mathbf{k}_1}(\mathbf{x}),\ldots,b_{N\mathbf{k}_1}(\mathbf{x}), \nonumber \\
& & b_{1\mathbf{k}_2}(\mathbf{x}),b_{2\mathbf{k}_2}(\mathbf{x}),\ldots,b_{N\mathbf{k}_2}(\mathbf{x}), \nonumber \\
& & \cdots ]^\dag \label{eqn: orb_b}
\end{eqnarray*}
where $\mathbf{k}_1,\mathbf{k}_2,\ldots$ are points in the first Brillouin zone, and $a_{i\mathbf{k}_j}(\mathbf{x})$ and $b_{i\mathbf{k}_j}(\mathbf{x})$ are the one-electron Bloch orbitals of $\Psi_A$ and $\Psi_B$, where the orbitals in each determinant are taken from the same orthonormal set. The corresponding orbital transformation~\cite{king1967corresponding} can be used to generalize for different orthonormal sets.  Since the overlaps between orbitals with different $\mathbf{k}$-points vanishes, the one-electron operators can be carried out per $\mathbf{k}$-point (i.e., block by block).  The matrix elements, however, for the two-electron operators are in general not block diagonal with respect to the $\mathbf{k}$-points.  In cases, where the two-electron matrix elements of the spin-orbitals have a double noncoincidence~\cite{king1967corresponding} the matrix elements are again block diagonal, otherwise the matrix elements can be represented as a sum of periodic Coulomb and exact exchange energies, where the Filon integration strategy~\cite{bylaska2020filon} can be used to fold in the first Brillouin zone integration present in the exact exchange energies.


To compare the energy states, $E_i$, between calculations with periodic and free space boundary conditions, it is convenient when calculating the $\bra{\Psi_A} H \ket{\Psi_A}$, $\bra{\Psi_B} H \ket{\Psi_B}$, $\bra{\Psi_A} H \ket{\Psi_B}$, and $\bra{\Psi_B} H \ket{\Psi_A}$ matrix elements to shift the Hamiltonian by a constant equal to the Ewald ion-ion energy, Eq.~\ref{eqn:ewald}, plus the charge correction ($\frac{Q^2 M}{2 r_s}$) for systems with periodic conditions, and a constant equal to the free space ion-ion energy for free-space boundary conditions. It should be noted that the constant shift does not affect energy differences.


\subsection{Matrix elements from the one-electron operators}
\label{sec:one-electron-matrix}
The $H_1$ operator for a periodic system written in reciprocal-space containing $N$-electrons per unit cell is
\begin{equation*}
    H_1 = \sum_{i=1}^{N} h(\mathbf{G}_{(i)}) 
\end{equation*}
where the $h(\mathbf{G})$ function/operator is
\begin{equation*}
    h(\mathbf{G}) = \frac{1}{2} \mathbf{G}^2 +  \sum_{I=1}^{N_\text{atoms}} \left(V^{I}_\text{local}({\bf G}) + \hat{V}^{I}_\text{NL} \right)
\end{equation*}
The periodic form of local pseudopotential is
given in Eq.~\ref{eqn:V_loc}, and based on Eq.~\ref{eqn:E_NL} the nonlocal pseudopotential kernel is defined as 
\begin{equation}\label{eqn:V_NL}
V_\text{NL}^I(\mathbf{G},\mathbf{G}') = \sum_{lm} \sum_{n,n'} P_{nlm}^{I}(\mathbf{G}) h_{l}^{n,n';I} P_{n'lm}^{I}(\mathbf{G}')
\end{equation}

The one-electron matrix elements between $\ket{\Psi_g}$, $\ket{\Psi_e}$, and $\ket{\Psi_m}$ states of the periodic $3\times3$ select CI Hamiltonian can be written using the corresponding orbitals formulas~\cite{bylaska2018corresponding} as the following.

\begin{eqnarray*}
    H_1^{gg} = \bra{\Psi_g} H_1 \ket{\Psi_g} &=& 2\sum_{i=1}^N\bigg(\frac{1}{2}\sum_{\mathbf{G}} \mathbf{G}^2 \psi_{f_i}^{*}(\mathbf{G}) \psi_{f_i}(\mathbf{G})  + \sum_{I,\mathbf{G}} V_\text{local}^I(\mathbf{G}) [\rho_{f_i, f_i}(\mathbf{G})]\\
    &+& \sum_I \sum_{\mathbf{G},\mathbf{G}'}  \psi_{f_i}^{*}(\mathbf{G}) V_\text{NL}^{I}(\mathbf{G},\mathbf{G}') \psi_{f_i}(\mathbf{G}')\bigg)
\end{eqnarray*}
\begin{eqnarray*}
    H_1^{ge} = \bra{\Psi_g} H_1 \ket{\Psi_e} &=& 0
\end{eqnarray*}
\begin{eqnarray*}   
    H_1^{gm} = \bra{\Psi_g} H_1 \ket{\Psi_m} &=& \sqrt{2}\bigg(\frac{1}{2}\sum_{\mathbf{G}} \mathbf{G}^2 \psi_{f_N}^{*}(\mathbf{G}) \psi_{e}(\mathbf{G})  + \sum_{I,\mathbf{G}} V_\text{local}^I(\mathbf{G}) [\rho_{f_N, e}(\mathbf{G})]\\  
    &+& \sum_I \sum_{\mathbf{G},\mathbf{G}'}  \psi_{f_N}^{*}(\mathbf{G}) V_\text{NL}^{I}(\mathbf{G},\mathbf{G}') \psi_{e}(\mathbf{G}')\bigg)
\end{eqnarray*}
\begin{eqnarray*}  
    H_1^{eg} = \bra{\Psi_e} H_1 \ket{\Psi_g} &=& 0
\end{eqnarray*}
\begin{eqnarray*}  
    H_1^{ee} = \bra{\Psi_e} H_1 \ket{\Psi_e} &=& 2\sum_{i=1}^{N-1}\bigg(\frac{1}{2}\sum_{\mathbf{G}} \mathbf{G}^2 \psi_{f_i}^{*}(\mathbf{G}) \psi_{f_i}(\mathbf{G})  + \sum_{I,\mathbf{G}} V_\text{local}^I(\mathbf{G}) [\rho_{f_i, f_i}(\mathbf{G})]\\  
    &+& \sum_I \sum_{\mathbf{G},\mathbf{G}'}  \psi_{f_i}^{*}(\mathbf{G}) V_\text{NL}^{I}(\mathbf{G},\mathbf{G}') \psi_{f_i}(\mathbf{G}')\bigg) + 2\bigg(\frac{1}{2}\sum_{\mathbf{G}} \mathbf{G}^2 \psi_{e}^{*}(\mathbf{G}) \psi_{e}(\mathbf{G})\\  
    &+& \sum_{I,\mathbf{G}} V_\text{local}^I(\mathbf{G}) [\rho_{e, e}(\mathbf{G})]  + \sum_I \sum_{\mathbf{G},\mathbf{G}'}  \psi_{e}^{*}(\mathbf{G}) V_\text{NL}^{I}(\mathbf{G},\mathbf{G}') \psi_{e}(\mathbf{G}')\bigg)
\end{eqnarray*}
\begin{eqnarray*}  
    H_1^{em} = \bra{\Psi_e} H_1 \ket{\Psi_m} &=& \sqrt{2}\bigg(\frac{1}{2}\sum_{\mathbf{G}} \mathbf{G}^2 \psi_{e}^{*}(\mathbf{G}) \psi_{f_N}(\mathbf{G})  + \sum_{I,\mathbf{G}} V_\text{local}^I(\mathbf{G}) [\rho_{e, f_N}(\mathbf{G})]\\
    &+& \sum_I \sum_{\mathbf{G},\mathbf{G}'}  \psi_{e}^{*}(\mathbf{G}) V_\text{NL}^{I}(\mathbf{G},\mathbf{G}') \psi_{f_N}(\mathbf{G}')\bigg)
\end{eqnarray*}
\begin{eqnarray*}  
    H_1^{mg}=\bra{\Psi_m} H_1 \ket{\Psi_g} &=& \sqrt{2}\bigg(\frac{1}{2}\sum_{\mathbf{G}} \mathbf{G}^2 \psi_{e}^{*}(\mathbf{G}) \psi_{f_N}(\mathbf{G})  + \sum_{I,\mathbf{G}} V_\text{local}^I(\mathbf{G}) [\rho_{e, f_N}(\mathbf{G})]\\
    &+& \sum_I \sum_{\mathbf{G},\mathbf{G}'}  \psi_{e}^{*}(\mathbf{G}) V_\text{NL}^{I}(\mathbf{G},\mathbf{G}') \psi_{f_N}(\mathbf{G}')\bigg)
\end{eqnarray*}
\begin{eqnarray*}  
    H_1^{me}=\bra{\Psi_m} H_1 \ket{\Psi_e} &=&\sqrt{2}\bigg(\frac{1}{2}\sum_{\mathbf{G}} \mathbf{G}^2 \psi_{f_N}^{*}(\mathbf{G}) \psi_{e}(\mathbf{G})  + \sum_{I,\mathbf{G}} V_\text{local}^I(\mathbf{G}) [\rho_{f_N, e}(\mathbf{G})]\\
    &+& \sum_I \sum_{\mathbf{G},\mathbf{G}'}  \psi_{f_N}^{*}(\mathbf{G}) V_\text{NL}^{I}(\mathbf{G},\mathbf{G}') \psi_{e}(\mathbf{G}')\bigg)
\end{eqnarray*}
\begin{eqnarray*}  
    H_1^{mm}= \bra{\Psi_m} H_1 \ket{\Psi_m} &=& H_1^{gg} + H_1^{ee}\\
    &=& 4\sum_{i=1}^{N-1}\bigg(\frac{1}{2}\sum_{\mathbf{G}} \mathbf{G}^2 \psi_{f_i}^{*}(\mathbf{G}) \psi_{f_i}(\mathbf{G})  + \sum_{I,\mathbf{G}} V_\text{local}^I(\mathbf{G}) [\rho_{f_i, f_i}(\mathbf{G})]\\
    &+& \sum_I \sum_{\mathbf{G},\mathbf{G}'}  \psi_{f_i}^{*}(\mathbf{G}) V_\text{NL}^{I}(\mathbf{G},\mathbf{G}') \psi_{f_i}(\mathbf{G}')\bigg)+2\bigg(\frac{1}{2}\sum_{\mathbf{G}} \mathbf{G}^2 \psi_{f_N}^{*}(\mathbf{G}) \psi_{f_N}(\mathbf{G})\\
    &+& \sum_{I,\mathbf{G}} V_\text{local}^I(\mathbf{G}) [\rho_{f_N, f_N}(\mathbf{G})] + \sum_I \sum_{\mathbf{G},\mathbf{G}'}  \psi_{f_N}^{*}(\mathbf{G}) V_\text{NL}^{I}(\mathbf{G},\mathbf{G}') \psi_{f_N}(\mathbf{G}')\bigg)\\
    &+& 2\bigg(\frac{1}{2}\sum_{\mathbf{G}} \mathbf{G}^2 \psi_{e}^{*}(\mathbf{G}) \psi_{e}(\mathbf{G})  + \sum_{I,\mathbf{G}} V_\text{local}^I(\mathbf{G}) [\rho_{e, e}(\mathbf{G})]\\  
    &+& \sum_I \sum_{\mathbf{G},\mathbf{G}'}  \psi_{e}^{*}(\mathbf{G}) V_\text{NL}^{I}(\mathbf{G},\mathbf{G}') \psi_{e}(\mathbf{G}')\bigg)
\end{eqnarray*}

The variation of these integrals with respect to $\psi_e^{*}(\mathbf{G})$ are then 

\begin{eqnarray*}
    \cfrac{\delta H_1^{gg}}{\delta \psi_e^{*}(\mathbf{G})} = \cfrac{\delta \bra{\Psi_g} H_1 \ket{\Psi_g}}{\delta \psi_e^{*}(\mathbf{G})} &=& 0
\end{eqnarray*}
\begin{eqnarray*}  
    \cfrac{\delta H_1^{ge}}{\delta \psi_e^{*}(\mathbf{G})} = \cfrac{\delta \bra{\Psi_g} H_1 \ket{\Psi_e}}{\delta \psi_e^{*}(\mathbf{G})} &=& 0
\end{eqnarray*}
\begin{eqnarray*}  
    \cfrac{\delta H_1^{gm}}{\delta \psi_e^{*}(\mathbf{G})} = \cfrac{\delta \bra{\Psi_g} H_1 \ket{\Psi_m}}{\delta \psi_e^{*}(\mathbf{G})} &=& 0
\end{eqnarray*}
\begin{eqnarray*}  
    \cfrac{\delta H_1^{eg}}{\delta \psi_e^{*}(\mathbf{G})} = \cfrac{\delta \bra{\Psi_e} H_1 \ket{\Psi_g}}{\delta \psi_e^{*}(\mathbf{G})} &=& 0
\end{eqnarray*}
\begin{eqnarray*}  
    \cfrac{\delta H_1^{ee}}{\delta \psi_e^{*}(\mathbf{G})} = \cfrac{\delta \bra{\Psi_e} H_1 \ket{\Psi_e}}{\delta \psi_e^{*}(\mathbf{G})} &=& 2\bigg(\frac{1}{2} \mathbf{G}^2  \psi_e(\mathbf{G}) +\sum_{I}\sum_{\mathbf{G}'} V_\text{local}^I(\mathbf{G}') \psi_{e}(\mathbf{G}+\mathbf{G}')\\
    &+& \sum_I \sum_{\mathbf{G}'}   V_\text{NL}^{I}(\mathbf{G},\mathbf{G}') \psi_e(\mathbf{G}')\bigg)
\end{eqnarray*}
\begin{eqnarray*}  
    \cfrac{\delta H_1^{em}}{\delta \psi_e^{*}(\mathbf{G})} = \cfrac{\delta \bra{\Psi_e} H_1 \ket{\Psi_m}}{\delta \psi_e^{*}(\mathbf{G})} &=&\sqrt{2}\bigg(\frac{1}{2} \mathbf{G}^2  \psi_{f_N}(\mathbf{G}) +\sum_{I}\sum_{\mathbf{G}'} V_\text{local}^I(\mathbf{G}') \psi_{f_N}(\mathbf{G}+\mathbf{G}')\\
    &+& \sum_I \sum_{\mathbf{G}'}   V_\text{NL}^{I}(\mathbf{G},\mathbf{G}') \psi_{f_N}(\mathbf{G}')\bigg)
\end{eqnarray*}
\begin{eqnarray*}  
    \cfrac{\delta H_1^{mg}}{\delta \psi_e^{*}(\mathbf{G})} = \cfrac{\delta \bra{\Psi_m} H_1 \ket{\Psi_g}}{\delta \psi_e^{*}(\mathbf{G})} &=& \sqrt{2}\bigg(\frac{1}{2} \mathbf{G}^2  \psi_{f_N}(\mathbf{G}) +\sum_{I}\sum_{\mathbf{G}'} V_\text{local}^I(\mathbf{G}') \psi_{f_N}(\mathbf{G}+\mathbf{G}')\\
    &+& \sum_I \sum_{\mathbf{G}'}   V_\text{NL}^{I}(\mathbf{G},\mathbf{G}') \psi_{f_N}(\mathbf{G}')\bigg)
\end{eqnarray*}
\begin{eqnarray*}  
    \cfrac{\delta H_1^{me}}{\delta \psi_e^{*}(\mathbf{G})} = \cfrac{\delta \bra{\Psi_m} H_1 \ket{\Psi_e}}{\delta \psi_e^{*}(\mathbf{G})} &=& 0
\end{eqnarray*}
\begin{eqnarray*}  
    \cfrac{\delta H_1^{mm}}{\delta \psi_e^{*}(\mathbf{G})} = \cfrac{\delta \bra{\Psi_m} H_1 \ket{\Psi_m}}{\delta \psi_e^{*}(\mathbf{G})} &=& 2\bigg(\frac{1}{2} \mathbf{G}^2  \psi_e(\mathbf{G}) +\sum_{I}\sum_{\mathbf{G}'} V_\text{local}^I(\mathbf{G}') \psi_{e}(\mathbf{G}+\mathbf{G}')\\
    &+& \sum_I \sum_{\mathbf{G}'}   V_\text{NL}^{I}(\mathbf{G},\mathbf{G}') \psi_e(\mathbf{G}')\bigg)
\end{eqnarray*}

where $V_\text{local}^I(\mathbf{G}')$ and $V_\text{NL}^{I}(\mathbf{G},\mathbf{G}')$ are given in Eq.~\ref{eqn:V_loc} and Eq.~\ref{eqn:V_NL}. 

\subsection{Matrix elements from the two-electron operators}
\label{sec:two-electron-matrix}


The two-electron matrix elements between $\ket{\Psi_g}$, $\ket{\Psi_e}$, and $\ket{\Psi_m}$ states of the periodic $3\times3$ select CI Hamiltonian can be written using the Slater-Condon rules or the corresponding orbitals formulas~\cite{bylaska2018corresponding,bylaska2020filon} as the following.

\begin{eqnarray*}
    H_2^{gg} &=& \bra{\Psi_g} H_2 \ket{\Psi_g} = E_\text{H}^{gg}+E_{\text{exch}}^{gg}\\
    &=& \frac{\Omega}{2} \sum_{i=1}^{N}\sum_{j=1}^{N}\sum_{\mathbf{G\neq0}} [\rho^{*}_{f_i, f_i}(\mathbf{G})] \frac{4\pi}{|\mathbf{G}|^2} [\rho_{f_j, f_j}(\mathbf{G})]-\frac{\Omega}{2} \sum_{i=1}^{N}\sum_{j=1}^{N}\sum_{\mathbf{G}}
[\rho^{*}_{f_i, f_j}(\mathbf{G})][\rho_{f_j, f_i}(\mathbf{G})]
V_{f}(\mathbf{G})
\end{eqnarray*}
\begin{eqnarray*}  
    H_2^{ge} &=& \bra{\Psi_g} H_2 \ket{\Psi_e} = E_{\text{screened}}^{ge}=\frac{\Omega}{2}\sum_{\mathbf{G}} [\rho^{*}_{f_N, e}(\mathbf{G})] [\rho_{f_N, e}(\mathbf{G})] V_{f}(\mathbf{G})
\end{eqnarray*}
\begin{eqnarray*}  
    H_2^{gm} &=& \bra{\Psi_g} H_2 \ket{\Psi_m} =
    E_\text{H}^{gm}+E_{\text{exch}}^{gm}\\
    &=&\sqrt{2}\bigg(\frac{\Omega}{2}\sum_{i=1}^{N}\sum_{\mathbf{G\neq0}} [\rho^{*}_{f_i, f_i}(\mathbf{G})] \frac{4\pi}{|\mathbf{G}|^2} [\rho_{f_N, e}(\mathbf{G})]-\frac{\Omega}{2}\sum_{i=1}^{N} \sum_{\mathbf{G}}[\rho^{*}_{f_i, e}(\mathbf{G})] [\rho_{f_N, f_i}(\mathbf{G})]V_{f}(\mathbf{G})\bigg)
\end{eqnarray*}
\begin{eqnarray*}  
    H_2^{eg} &=& \bra{\Psi_e} H_2 \ket{\Psi_g} =
    E_{\text{screened}}^{eg}=\frac{\Omega}{2}\sum_{\mathbf{G}}[\rho^{*}_{e, f_N}(\mathbf{G})] [\rho_{e, f_N}(\mathbf{G})]
V_{f}(\mathbf{G})
\end{eqnarray*}
\begin{eqnarray*}  
    H_2^{ee} &=& \bra{\Psi_e} H_2 \ket{\Psi_e} =
    E_\text{H}^{ee}+E_{\text{exch}}^{ee}\\
    &=& \frac{\Omega}{2} \sum_{i=1}^{N-1}\sum_{j=1}^{N-1}\sum_{\mathbf{G\neq0}} [\rho^{*}_{f_i, f_i}(\mathbf{G})] \frac{4\pi}{|\mathbf{G}|^2} [\rho_{f_j, f_j}(\mathbf{G})] + \frac{\Omega}{2}\sum_{i=1}^{N-1}\sum_{\mathbf{G\neq0}} [\rho^{*}_{f_i, f_i}(\mathbf{G})] \frac{4\pi}{|\mathbf{G}|^2} [\rho_{e, e}(\mathbf{G})]\\
    &+&\frac{\Omega}{2}\sum_{\mathbf{G\neq0}} [\rho^{*}_{e, e}(\mathbf{G})] \frac{4\pi}{|\mathbf{G}|^2} [\rho_{e, e}(\mathbf{G})]-\frac{\Omega}{2} \sum_{i=1}^{N-1}\sum_{j=1}^{N-1}\sum_{\mathbf{G}}
[\rho^{*}_{f_i, f_j}(\mathbf{G})] [\rho_{f_j, f_i}(\mathbf{G})]
V_{f}(\mathbf{G})\\
&-&\frac{\Omega}{2} \sum_{i=1}^{N-1}\sum_{\mathbf{G}}
[\rho^{*}_{f_i, e}(\mathbf{G})] [\rho_{e, f_i}(\mathbf{G})]
V_{f}(\mathbf{G})-\frac{\Omega}{2}\sum_{\mathbf{G}}
[\rho^{*}_{e, e}(\mathbf{G})] [\rho_{e, e}(\mathbf{G})]
V_{f}(\mathbf{G})
\end{eqnarray*}
\begin{eqnarray*}  
    H_2^{em} &=& \bra{\Psi_e} H_2 \ket{\Psi_m} = E_\text{H}^{em}+E_{\text{exch}}^{em}\\
    &=& \sqrt{2}\bigg(\frac{\Omega}{2}\sum_{i=1}^{N-1}\sum_{\mathbf{G\neq0}} [\rho^{*}_{f_i, f_i}(\mathbf{G})] \frac{4\pi}{|\mathbf{G}|^2} [\rho_{e, f_N}(\mathbf{G})] +\frac{\Omega}{2}\sum_{\mathbf{G\neq0}} [\rho^{*}_{e, f_N}(\mathbf{G})] \frac{4\pi}{|\mathbf{G}|^2} [\rho_{e, e}(\mathbf{G})]\\
    &-&\frac{\Omega}{2} \sum_{i=1}^{N-1}\sum_{\mathbf{G}}
[\rho^{*}_{f_i, f_N}(\mathbf{G})] [\rho_{e, f_i}(\mathbf{G})]
V_{f}(\mathbf{G})-\frac{\Omega}{2}\sum_{\mathbf{G}}
[\rho^{*}_{e, e}(\mathbf{G})] [\rho_{e, f_N}(\mathbf{G})]
V_{f}(\mathbf{G})\bigg)
\end{eqnarray*}
\begin{eqnarray*}  
    H_2^{mg} &=& \bra{\Psi_m} H_2 \ket{\Psi_g} = 
    E_\text{H}^{mg}+E_{\text{exch}}^{mg}\\
    &=& \sqrt{2}\bigg(\frac{\Omega}{2}\sum_{i=1}^{N}\sum_{\mathbf{G\neq0}} [\rho^{*}_{f_i, f_i}(\mathbf{G})] \frac{4\pi}{|\mathbf{G}|^2} [\rho_{e, f_N}(\mathbf{G})]-\frac{\Omega}{2} \sum_{i=1}^{N}\sum_{\mathbf{G}}[\rho^{*}_{f_i, f_N}(\mathbf{G})] [\rho_{e, f_i}(\mathbf{G})]V_{f}(\mathbf{G})\bigg)
\end{eqnarray*}
\begin{eqnarray*}  
    H_2^{me} &=& \bra{\Psi_m} H_2 \ket{\Psi_e} = E_\text{H}^{me}+E_{\text{exch}}^{me}\\
    &=& \sqrt{2}\bigg(\frac{\Omega}{2}\sum_{i=1}^{N-1}\sum_{\mathbf{G\neq0}} [\rho^{*}_{f_i, f_i}(\mathbf{G})] \frac{4\pi}{|\mathbf{G}|^2} [\rho_{f_N, e}(\mathbf{G})]+\frac{\Omega}{2}\sum_{\mathbf{G\neq0}} [\rho^{*}_{f_N, e}(\mathbf{G})] \frac{4\pi}{|\mathbf{G}|^2} [\rho_{e, e}(\mathbf{G})]\\
    &-&\frac{\Omega}{2} \sum_{i=1}^{N-1}\sum_{\mathbf{G}} [\rho^{*}_{f_i, e}(\mathbf{G})] [\rho_{f_N, f_i}(\mathbf{G})]V_{f}(\mathbf{G})-\frac{\Omega}{2}\sum_{\mathbf{G}}[\rho^{*}_{f_N, e}(\mathbf{G})] [\rho_{e, e}(\mathbf{G})]
V_{f}(\mathbf{G})\bigg)
\end{eqnarray*}
\begin{eqnarray*}
H_2^{mm} &=& \bra{\Psi_m} H_2 \ket{\Psi_m} = E_\text{H}^{mm}+E_{\text{exch}}^{mm}\\
    &=& \frac{\Omega}{2}\sum_{i=1}^{N-1}\sum_{j=1}^{N-1}\sum_{\mathbf{G\neq0}} [\rho_{f_i, f_i}(\mathbf{-G})] \frac{4\pi}{|\mathbf{G}|^2} [\rho_{f_j, f_j}(\mathbf{G})] +\frac{\Omega}{2}\sum_{i=1}^{N-1}\sum_{\mathbf{G\neq0}} [\rho^{*}_{f_i, f_i}(\mathbf{G})] \frac{4\pi}{|\mathbf{G}|^2} [\rho_{f_N, f_N}(\mathbf{G})]\\
    &+&\frac{\Omega}{2}\sum_{i=1}^{N}\sum_{\mathbf{G\neq0}} [\rho^{*}_{f_i, f_i}(\mathbf{G})] \frac{4\pi}{|\mathbf{G}|^2} [\rho_{e, e}(\mathbf{G})]-\frac{\Omega}{2} \sum_{i=1}^{N-1}\sum_{j=1}^{N-1}\sum_{\mathbf{G}}[\rho^{*}_{f_i, f_j}(\mathbf{G})] [\rho_{f_j, f_i}(\mathbf{G})]V_{f}(\mathbf{G})\\
    &-&\frac{\Omega}{2} \sum_{i=1}^{N-1}\sum_{\mathbf{G}}[\rho^{*}_{f_i, f_N}(\mathbf{G})] [\rho_{f_N, f_i}(\mathbf{G})]
V_{f}(\mathbf{G})-\frac{\Omega}{2} \sum_{i=1}^{N}\sum_{\mathbf{G}}
[\rho^{*}_{f_i, e}(\mathbf{G})] [\rho_{e, f_i}(\mathbf{G})]
V_{f}(\mathbf{G})\\
&+&\frac{\Omega}{2}\sum_{\mathbf{G}}
[\rho^{*}_{f_N, e}(\mathbf{G})] [\rho_{e, f_N}(\mathbf{G})]
V_{f}(\mathbf{G})
\end{eqnarray*}

The variation of the two-electron integrals with respect to $\psi_e^{*}(\mathbf{G})$ are then 

\begin{eqnarray*}
    \cfrac{\delta H_2^{gg}}{\delta \psi_e^{*}(\mathbf{G})}=\cfrac{\delta \bra{\Psi_g} H_2 \ket{\Psi_g}}{\delta \psi_e^{*}(\mathbf{G})} &=& 0 
\end{eqnarray*}
\begin{eqnarray*}  
    \cfrac{\delta H_2^{ge}}{\delta \psi_e^{*}(\mathbf{G})}=\cfrac{\delta \bra{\Psi_g} H_2 \ket{\Psi_e}}{\delta \psi_e^{*}(\mathbf{G})} &=& 0 
\end{eqnarray*}
\begin{eqnarray*}  
    \cfrac{\delta H_2^{gm}}{\delta \psi_e^{*}(\mathbf{G})}=\cfrac{\delta \bra{\Psi_g} H_2 \ket{\Psi_m}}{\delta \psi_e^{*}(\mathbf{G})} &=& 0 
\end{eqnarray*}
\begin{eqnarray*}  
    \cfrac{\delta H_2^{eg}}{\delta \psi_e^{*}(\mathbf{G})}=\cfrac{\delta \bra{\Psi_e} H_2 \ket{\Psi_g}}{\delta \psi_e^{*}(\mathbf{G})} &=& \frac{\Omega}{2} \sum_{\mathbf{G}'} \psi_{f_N}(\mathbf{G}- \mathbf{G}')[\rho_{e, f_N}(\mathbf{G}')]V_{f}(\mathbf{G}')\\
    &+& \frac{\Omega}{2} \sum_{\mathbf{G}'} [\rho^{*}_{e, f_N}(\mathbf{G}')]\psi_{f_N}(\mathbf{G}+ \mathbf{G}')V_{f}(\mathbf{G}')
\end{eqnarray*}
\begin{eqnarray*}  
\cfrac{\delta H_2^{ee}}{\delta \psi_e^{*}(\mathbf{G})}=\cfrac{\delta \bra{\Psi_e} H_2 \ket{\Psi_e}}{\delta \psi_e^{*}(\mathbf{G})} &=& 
    \frac{\Omega}{2}\sum_{i=1}^{N-1}\sum_{\mathbf{G'\neq0}} [\rho^{*}_{f_i, f_i}(\mathbf{G}')] \frac{4\pi}{|\mathbf{G}|^2} \psi_{e}(\mathbf{G}+\mathbf{G}')\\
    &+& \frac{\Omega}{2}\sum_{\mathbf{G'\neq0}} \psi_{e}(\mathbf{G}-\mathbf{G}') \frac{4\pi}{|\mathbf{G}'|^2}[\rho_{e, e}(\mathbf{G}')] \\
    &+& \frac{\Omega}{2}\sum_{\mathbf{G'\neq0}} [\rho^{*}_{e, e}(\mathbf{G}')] \frac{4\pi}{|\mathbf{G}'|^2} \psi_{e}(\mathbf{G}+\mathbf{G}')\\
&-&\frac{\Omega}{2} \sum_{i=1}^{N-1}\sum_{\mathbf{G}'}
[\rho^{*}_{f_i, e}(\mathbf{G}')] \psi_{f_i}(\mathbf{G}+\mathbf{G}')V_{f}(\mathbf{G}')\\
&-&\frac{\Omega}{2}\sum_{\mathbf{G}'}\psi_{e}(\mathbf{G}-\mathbf{G}')[\rho_{e, e}(\mathbf{G}')]V_{f}(\mathbf{G}')\\
&-&\frac{\Omega}{2}\sum_{\mathbf{G}'}
[\rho^{*}_{e, e}(\mathbf{G}')] \psi_{e}(\mathbf{G}+\mathbf{G}')V_{f}(\mathbf{G}')
\end{eqnarray*}
\begin{eqnarray*}  
\cfrac{\delta H_2^{em}}{\delta \psi_e^{*}(\mathbf{G})}=\cfrac{\delta \bra{\Psi_e} H_2 \ket{\Psi_m}}{\delta \psi_e^{*}(\mathbf{G})} &=& 
    \sqrt{2}\bigg(\frac{\Omega}{2}\sum_{i=1}^{N-1}\sum_{\mathbf{G'\neq0}} [\rho^{*}_{f_i, f_i}(\mathbf{G}')] \frac{4\pi}{|\mathbf{G}'|^2} \psi_{f_N}(\mathbf{G}+\mathbf{G}')\\
    &+& \frac{\Omega}{2}\sum_{\mathbf{G'\neq0}} \psi_{f_N}(\mathbf{G}-\mathbf{G}') \frac{4\pi}{|\mathbf{G}'|^2} [\rho_{e, e}(\mathbf{G}')]\\
    &+& \frac{\Omega}{2}\sum_{\mathbf{G'\neq0}} [\rho^{*}_{e, f_N}(\mathbf{G}')] \frac{4\pi}{|\mathbf{G}'|^2} \psi_{e}(\mathbf{G}+\mathbf{G'})\\
    &-& \frac{\Omega}{2} \sum_{i=1}^{N-1}\sum_{\mathbf{G}'}
[\rho^{*}_{f_i, f_N}(\mathbf{G}')] \psi_{f_i}(\mathbf{G}+\mathbf{G}')V_{f}(\mathbf{G}')\\
&-&\frac{\Omega}{2}\sum_{\mathbf{G}'}
\psi_{e}(\mathbf{G}-\mathbf{G}')[\rho_{e, f_N}(\mathbf{G}')]V_{f}(\mathbf{G}')\\
&-& \frac{\Omega}{2}\sum_{\mathbf{G}'}
[\rho^{*}_{e, e}(\mathbf{G}')] \psi_{f_N}(\mathbf{G}+\mathbf{G}')V_{f}(\mathbf{G}')\bigg)
\end{eqnarray*}
\begin{eqnarray*}  
    \cfrac{\delta H_2^{mg}}{\delta \psi_e^{*}(\mathbf{G})}=\cfrac{\delta \bra{\Psi_m} H_2 \ket{\Psi_g}}{\delta \psi_e^{*}(\mathbf{G})} &=& \sqrt{2}\bigg(\frac{\Omega}{2}\sum_{i=1}^{N}\sum_{\mathbf{G'\neq0}} [\rho^{*}_{f_i, f_i}(\mathbf{G}')] \frac{4\pi}{|\mathbf{G}'|^2} \psi_{f_N}(\mathbf{G}+\mathbf{G}')\\
    &-& \frac{\Omega}{2}\sum_{i=1}^{N} \sum_{\mathbf{G}'}\rho^{*}_{f_i f_N}(\mathbf{G}') \psi_{f_i}(\mathbf{G}+\mathbf{G}')V_{f}(\mathbf{G}')\bigg)
\end{eqnarray*}
\begin{eqnarray*}  
    \cfrac{\delta H_2^{me}}{\delta \psi_e^{*}(\mathbf{G})}=\cfrac{\delta \bra{\Psi_m} H_2 \ket{\Psi_e}}{\delta \psi_e^{*}(\mathbf{G})} &=& \sqrt{2}\bigg(\frac{\Omega}{2}\sum_{\mathbf{G'\neq0}} [\rho^{*}_{f_N, e}(\mathbf{G}')] \frac{4\pi}{|\mathbf{G}'|^2} \psi_{e}(\mathbf{G}+\mathbf{G}')\\
    &-&\frac{\Omega}{2}\sum_{\mathbf{G}'}
[\rho^{*}_{f_N, e}(\mathbf{G}')] \psi_{e}(\mathbf{G}+\mathbf{G}')V_{f}(\mathbf{G}')\bigg)
\end{eqnarray*}
\begin{eqnarray*}
\cfrac{\delta H_2^{mm}}{\delta \psi_e^{*}(\mathbf{G})}=\cfrac{\delta \bra{\Psi_m} H_2 \ket{\Psi_m}}{\delta \psi_e^{*}(\mathbf{G})} &=&
    \frac{\Omega}{2}\sum_{i=1}^{N}\sum_{\mathbf{G'\neq0}} [\rho^{*}_{f_i, f_i}(\mathbf{G}')] \frac{4\pi}{|\mathbf{G}'|^2} \psi_{e }(\mathbf{G}+\mathbf{G}')\\
    &-& \frac{\Omega}{2} \sum_{i=1}^{N}\sum_{\mathbf{G}'}
[\rho^{*}_{f_i, e}(\mathbf{G}')] \psi_{f_i}(\mathbf{G}+\mathbf{G}')V_{f}(\mathbf{G}')\\
&+&\frac{\Omega}{2}\sum_{\mathbf{G}'}
[\rho^{*}_{f_N, e}(\mathbf{G}')] \psi_{f_N}(\mathbf{G}+\mathbf{G}')V_{f}(\mathbf{G}')
\end{eqnarray*}

\section{Results for the Ground State of the LiH Molecule Using Periodic Boundary Conditions}
\label{sec:LiHresults}

The NWChem program package~\cite{kendall2000high,valiev2010nwchem,bylaska2011large,bylaska2017plane,apra2020nwchem} was used for all calculations in this study, except for the FCI calculations, which used the TINYMRCC suite by Ji{\v{r}}{\'\i} Pittner.  The plane-wave calculations used a simple cubic box with $L=15$ \AA, and a cutoff energy of 40 Ry. The web application EMSL Arrows~\cite{bylaska2021building} was used to set up and perform the plane-wave calculations. The valence electron interactions with the atomic core are approximated with generalized norm-conserving Hamann~\cite{hamann1989generalized} pseudopotentials modified to the separable form suggested by Kleinman and Bylander~\cite{kleinman1982efficacious}.  The pseudopotentials used in this study were constructed using the following core radii: H: rcs=0.800 a.u and rcp=0.800 a.u.; Li: rcs=1.869 a.u, and rcp=1.551 a.u..  

The results for PW FCI calculations of LiH with 1, 4, 8, 12, and 18 COVOs are shown in Figure~\ref{fig:lih-pw-fci} and Table~\ref{tab:lihpwfci} in Appendix~\ref{sec:appendix}.  Our code produced the whole energy curves that show inversion symmetry about the central point at $R=7.5$ \AA, i.e., the energy at the distance ($15$ \AA$-R$) produced the same energy found at $R$ with the simple-cubic supercell size of $15$ {\AA} due to the periodicity.  The average difference error for the 1, 4, 8, and 12 COVOs calculations  from the 18 COVOs calculation is 12.9 kcal/mol, 2.7 kcal/mol, 1.0 kcal/mol, and 0.4 kcal/mol respectively. While the error is significant for 1 virtual orbital, the difference is quite small by 4 virtual orbitals, and the error steadily decreases as the number of virtual orbitals is increased.  Another measure of the error is the extensivity error. The energy for large $R$ should be the same as the combined energy of the isolated H and Li atoms. The aperiodic PW FCI energies for the dissociated atoms (at {$R=7$ \mbox{\normalfont\AA}}) were found to be -0.66372, -0.68739, -0.68945, -0.68946, and -0.69011 Hartrees for 1, 4, 8, 12, and 18 optimized virtual orbitals, respectively. The sequence of numbers shows the convergence to the combined Hartree-Fock energy of the isolated H and Li atoms which is -0.691388 Hartrees ($E(\text{H})=-0.498883$ Hartrees and $E(\text{Li})=-0.192505$ Hartrees) calculated by the pseudopotential plane-wave method.

In Figure~\ref{fig:zoom-lih-pw-fci-compare2} we compared the total energies from aperiodic (see Table~\ref{tab:lihpwfci-apd} in Appendix~\ref{sec:appendix}) and periodic plane-wave FCI calculations for the LiH molecule with 1 and 18 correlation optimized virtual orbitals. The energies from periodic plane-wave FCI calculations are lower than the energies from aperiodic calculations from $R=1.3$ \AA \, to $R=3.5$ \AA, while the former are higher than the latter from $R=4.0$ \AA \, to $R=7.0$ \AA. The average difference error for the 1 and 18 COVOs calculations between the aperiodic and periodic energies is 1.2 kcal/mol and 1.3 kcal/mol respectively, which suggests that periodic results agree with the aperiodic ones. However, at large $R$ a significant difference between aperiodic and periodic calculations can be observed. The comparison between the total energies from aperiodic and periodic plane-wave FCI calculations for the H$_2$ molecule with 8 correlation optimized virtual orbitals is shown in Figure~\ref{fig:zoom-h2-pw-fci-compare2}. The difference in the agreement between periodic and aperiodic energies at large $R$ for LiH and H$_2$ molecules is due to the dipoles in molecules. Since H$_2$ is a non-polar molecule, there are no dipoles that affect the total energies in both aperiodic and periodic systems while for the polar LiH molecule at large $R$, the dipoles between Li and H atoms and their images in periodic systems cancel each other in the periodic systems and thus the energy becomes higher than the energies in the aperiodic system.

\begin{figure}[htp]
    \centering
    \begin{tikzpicture}[scale=1.35]
    \begin{axis}[my style,grid=major,minor tick num=1,xmin=1.0,xmax=14.0,ymin=-0.785,ymax=-0.59,legend style={nodes={scale=0.6}},legend pos=north east] 
        \addplot [smooth,thin,red]    table [x=R(Li-H), y=1 COVO,  col sep=comma] {data/LiH-periodic.csv};
        \addplot [smooth,thin,orange] table [x=R(Li-H), y=4 COVOs, col sep=comma] {data/LiH-periodic.csv};
        \addplot [smooth,thin,green]  table [x=R(Li-H), y=8 COVOs, col sep=comma] {data/LiH-periodic.csv};
        \addplot [smooth,thin,blue]   table [x=R(Li-H), y=12 COVOs,col sep=comma] {data/LiH-periodic.csv};
        \addplot [smooth,thin,black]  table [x=R(Li-H), y=18 COVOs,col sep=comma] {data/LiH-periodic.csv};
        \legend{PW FCI - 1 COVO, PW FCI - 4 COVOs, PW FCI - 8 COVOs, PW FCI - 12 COVOs, PW FCI - 18 COVOs}
    \end{axis}
    \end{tikzpicture}  
    \begin{tikzpicture}[scale=1.35]
    \begin{axis}[my style,grid=major,minor tick num=1,xmin=1.3,xmax=2.5,ymin=-0.785,ymax=-0.710, legend pos=north east, legend style={nodes={scale=0.6}}]
        \addplot [smooth,thick,red,mark=* ]   table [x=R(Li-H), y=1 COVO,  col sep=comma] {data/LiH-periodic.csv};
        \addplot [smooth,thick,orange,mark=*] table [x=R(Li-H), y=4 COVOs, col sep=comma] {data/LiH-periodic.csv};
        \addplot [smooth,thick,green,mark=*]  table [x=R(Li-H), y=8 COVOs, col sep=comma] {data/LiH-periodic.csv};
        \addplot [smooth,thick,blue,mark=*]   table [x=R(Li-H), y=12 COVOs,col sep=comma] {data/LiH-periodic.csv};
        \addplot [smooth,thick,black,mark=*]  table [x=R(Li-H), y=18 COVOs,col sep=comma] {data/LiH-periodic.csv};
        \legend{PW FCI - 1 COVO, PW FCI - 4 COVOs, PW FCI - 8 COVOs, PW FCI - 12 COVOs, PW FCI - 18 COVOs}
    \end{axis}
    \end{tikzpicture}  
    \caption{Total energies as a function of distance from periodic plane-wave FCI calculations for the LiH molecule with 1, 4, 8, 12, and 18 correlation optimized virtual orbitals. The top plot shows energy from $R$=1.3 \AA \, to $R$=13.7 \AA, and the bottom plot zooms in near the energy minima at $R$=1.6 \AA.  The periodic calculations used a simple-cubic supercell (L=15.0 \AA).}
	\label{fig:lih-pw-fci}
\end{figure}
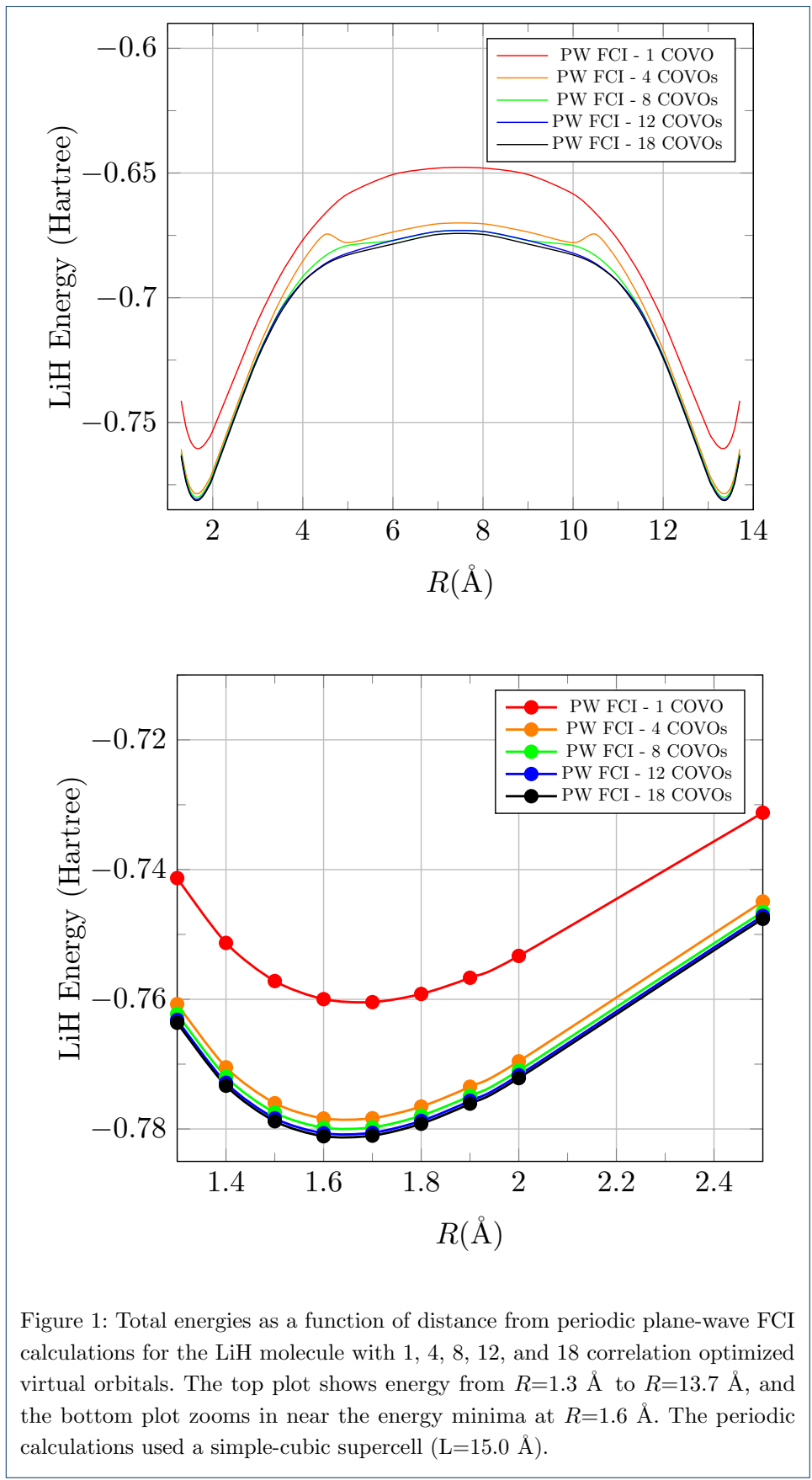

\begin{figure}[htp]
    \centering
        \begin{tikzpicture}[scale=1.35]
    \begin{axis}[my style,grid=major,minor tick num=1,xmin=1.3,xmax=2.5,ymin=-0.79,ymax=-0.71,legend pos=north east, legend style={nodes={scale=0.6}}]
        \addplot [smooth,thick,red,mark=* ]   table [x=R(Li-H), y=1 COVO,  col sep=comma] {data/LiH-periodic.csv};
        \addplot [smooth,thick,red,dashed] table [x=R(Li-H), y=1 COVO, col sep=comma] {data/LiH-aperiodic.csv};
        \addplot [smooth,thick,black,mark=*]  table [x=R(Li-H), y=18 COVOs,col sep=comma] {data/LiH-periodic.csv};
        \addplot [smooth,thick,dashed,black]  table [x=R(Li-H), y=18 COVOs,col sep=comma] {data/LiH-aperiodic.csv};
        \legend{Periodic - 1 COVO, Aperiodic - 1 COVO, Periodic - 18 COVOs, Aperiodic - 18 COVOs}
    \end{axis}
    \end{tikzpicture}  
        \begin{tikzpicture}[scale=1.35]
    \begin{axis}[my style,grid=major,minor tick num=1,xmin=2.5,xmax=7.0,ymin=-0.75,ymax=-0.60, legend pos=north east, legend style={nodes={scale=0.6}}]
        \addplot [smooth,thick,red,mark=* ]   table [x=R(Li-H), y=1 COVO,  col sep=comma] {data/LiH-periodic.csv};
        \addplot [smooth,thick,red,dashed] table [x=R(Li-H), y=1 COVO, col sep=comma] {data/LiH-aperiodic.csv};
        \addplot [smooth,thick,black,mark=*]  table [x=R(Li-H), y=18 COVOs,col sep=comma] {data/LiH-periodic.csv};
        \addplot [smooth,thick,dashed,black]  table [x=R(Li-H), y=18 COVOs,col sep=comma] {data/LiH-aperiodic.csv};
        \legend{Periodic - 1 COVO, Aperiodic - 1 COVO, Periodic - 18 COVOs, Aperiodic - 18 COVOs}
    \end{axis}
    \end{tikzpicture}  
    \caption{Total energies as a function of distance from aperiodic and periodic plane-wave FCI calculations for the LiH molecule with 1 and 18 correlation optimized virtual orbitals. The top plot shows energy from $R$=1.3 \AA \, to $R$=2.5 \AA, and the bottom plot shows energy from $R$=2.5 \AA \, to $R$=7.0 \AA.  The periodic calculations used a simple-cubic supercell (L=15.0 \AA).}
	\label{fig:zoom-lih-pw-fci-compare2}
\end{figure}

\begin{figure}[htp]
    \centering
        \begin{tikzpicture}[scale=1.35]
    \begin{axis}[H2 style,grid=major,minor tick num=1,xmin=0.6,xmax=1.5,ymin=-1.18,ymax=-1.04,legend pos=north east, legend style={nodes={scale=0.6}}]
        \addplot [smooth,thick,black,mark=*]  table [x=R(H-H), y=8 COVOs,col sep=comma] {data/H2-periodic.csv};
        \addplot [smooth,thick,dashed,yellow,mark=diamond*]  table [x=R(H-H), y=8 COVOs, col sep=comma] {data/H2-aperiodic.csv};
        \legend{Periodic - 8 COVOs, Aperiodic - 8 COVOs}
    \end{axis}
    \end{tikzpicture}  
        \begin{tikzpicture}[scale=1.35]
    \begin{axis}[H2 style,grid=major,minor tick num=1,xmin=1.5,xmax=12.26,ymin=-1.08,ymax=-0.97,legend pos=north east, legend style={nodes={scale=0.6}}]
        \addplot [smooth,thick,black,mark=*]  table [x=R(H-H), y=8 COVOs,col sep=comma] {data/H2-periodic.csv};
        \addplot [smooth,thick,dashed,yellow,mark=diamond*]  table [x=R(H-H), y=8 COVOs,col sep=comma] {data/H2-aperiodic.csv};
        \legend{Periodic - 8 COVOs, Aperiodic - 8 COVOs}
    \end{axis}
    \end{tikzpicture}  
    \caption{Total energies as a function of distance from aperiodic and periodic plane-wave FCI calculations for the H$_2$ molecule with 8 correlation optimized virtual orbitals. The top plot shows energy from $R$=0.6 \AA \, to $R$=1.5 \AA, and the bottom plot shows energy from $R$=1.5 \AA \, to $R$=12.3 \AA.}
	\label{fig:zoom-h2-pw-fci-compare2}
\end{figure}

\begin{figure}[htp]
    \centering
	\includegraphics[scale=0.35]{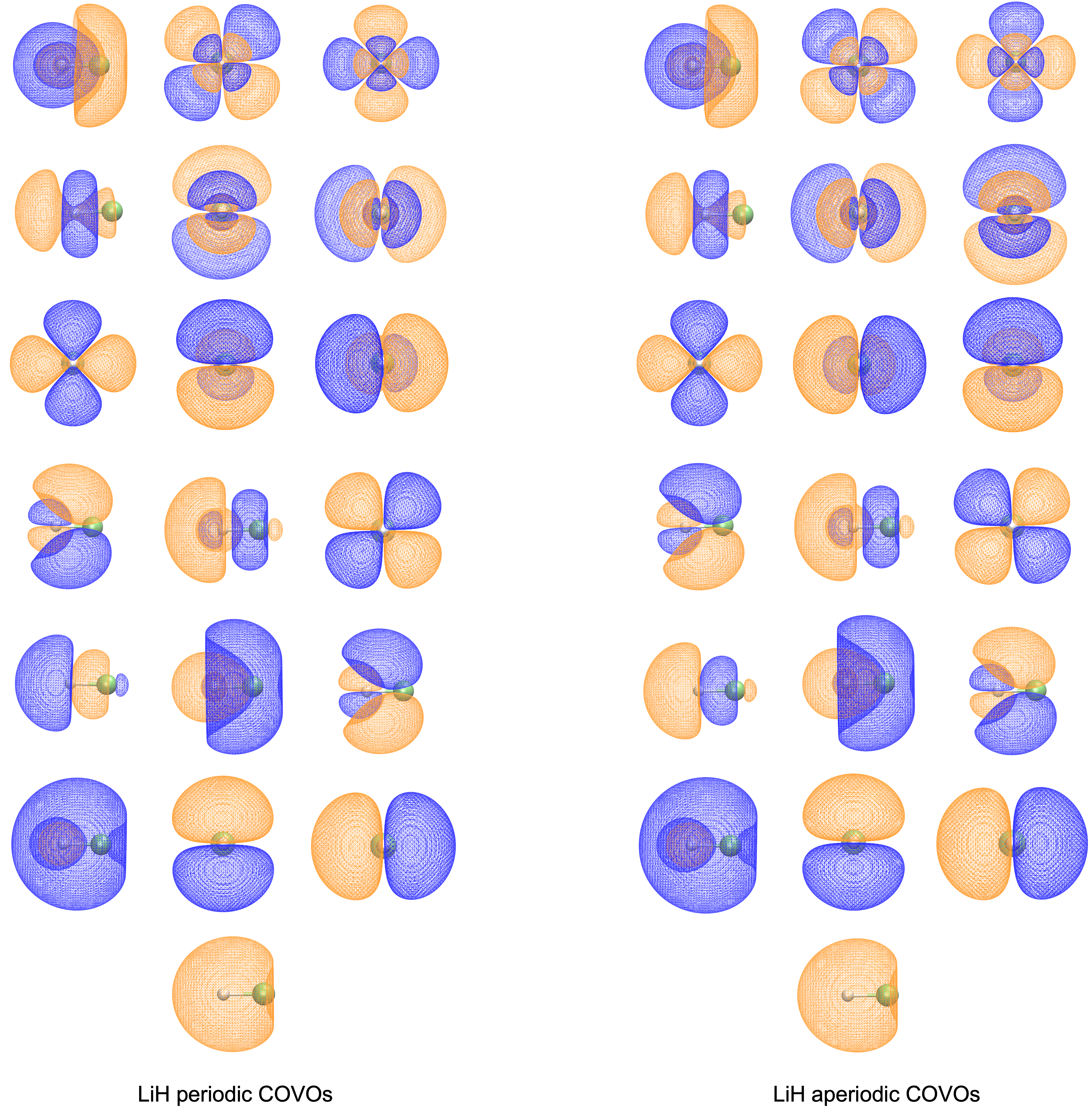}
	\caption{The 1 filled RHF orbital and 18 COVOs for the LiH molecule from periodic and aperiodic plane-wave FCI calculations are shown on the left and right panels respectively. The orbitals are displayed in the order of increasing orbital energy from left to right and bottom to top. The distance between two atoms at which the energy achieves its minimum is 1.6 \AA ~for LiH. The positive and negative isosurfaces are colored in blue and orange respectively.}
	\label{fig:LiH-periodic-aperiodic-covos}
\end{figure}

\vspace{1cm}
\section{Quantum Computer Calculations for the Ground State of the LiH Molecule Using Periodic Boundary Conditions}
\label{sec:LiHQCresults}

The COVOs optimized orbital basis sets can reduce the circuit depth and complexity for quantum algorithms, opening up applications in chemistry and physics. This is particularly meaningful as current and near-future quantum computers are noisy intermediate-scale quantum (NISQ) devices. These devices are restricted in the number of qubits, qubit connectivity, and fidelity of single- and multi-qubit entangling gates. To effectively utilize such quantum hardware, one must employ algorithms that minimize gate count and withstand noise, which is any undesired internal or external factor that changes the quantum system. Thanks to a grant from Microsoft, we were provided credits for jobs submitted on quantum computers and their corresponding emulators to demonstrate the applicability of the COVOs method in quantum computing for the periodic LiH system.

In this project, we had a few goals in mind in our endeavor to illustrate the applicability of a generated COVOs basis in quantum computing. First, we wanted to compute ground-state energy on an actual quantum computer. We also wanted ground-state energies at various Li--H internuclear distances to measure the performance of a quantum algorithm in different regimes or electronic correlation. Through Microsoft's Azure Quantum cloud computing service~\cite{Azure}, we had access to the H1-1 quantum computer provided by Quantinuum (a company formed from the combination of Honeywell Quantum Solutions and Cambridge Quantum)~\cite{Quantinuum}. The H1-1 quantum computer is the latest hardware in Quantinuum's H1 generation of quantum computers with high-fidelity, fully connected qubits. The high fidelity of qubits corresponds to lower errors brought on by noise. In addition to computing ground-state energies at different bond lengths on a quantum computer, we also wanted to measure the effects of noise on the corresponding circuit evaluations. With these goals in mind, we set out on this venture while being mindful of the available computational funds granted to us, which was achieved through a three-stage process. 

\begin{figure}[htp]
    \centering
	\includegraphics[scale=0.9]{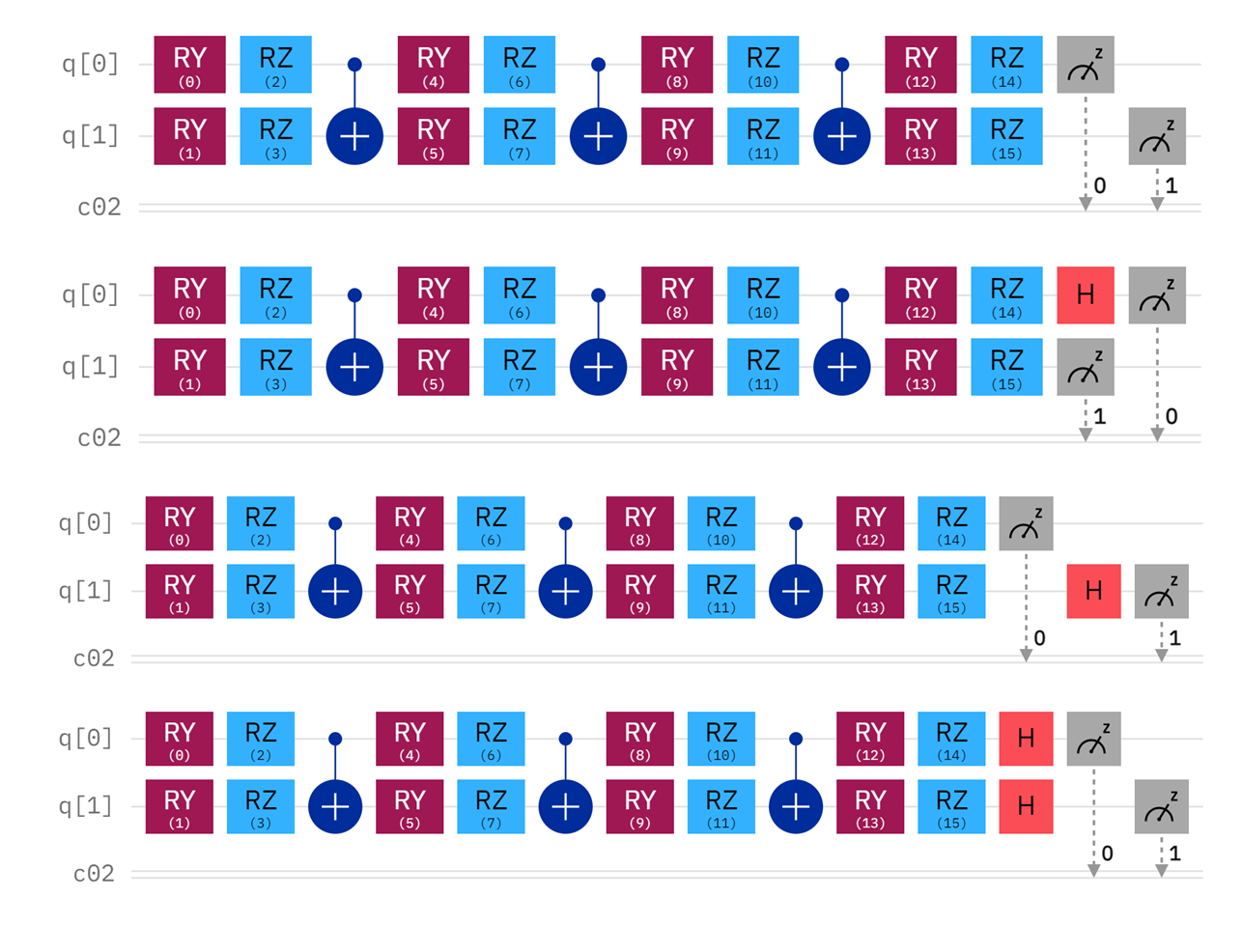}
	\caption{The two-level circuits consisting of 8 $R_y$ and 8 $R_y$ rotation gates, 3 Controlled-X (or Controlled-NOT) gates, and 2 measurements on the two qubits, along with either 0 (top), 1 (middle two), or 2 (bottom) Hadamard gates. The values in parentheses represent the 16 variational parameters in the circuit.}
	\label{fig:Circuit}
\end{figure}

\begin{figure}[htp]
    \centering
	\includegraphics[scale=1]{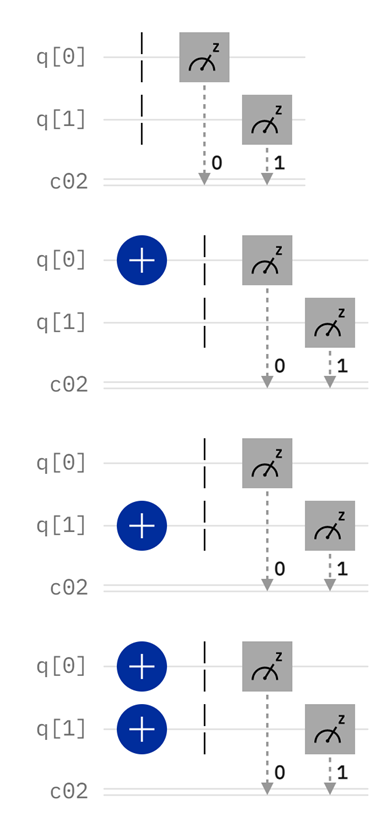}
	\caption{The four circuits used to construct the calibration matrix for the error mitigation.}
	\label{fig:calibration}
\end{figure}

We probed the potential energy surface of LiH at 1.7, 3.0, and 7.0 \mbox{\normalfont\AA} inter-nuclear distances employing VQE, one of the most widely used near-term applications for quantum computing that has successfully been deployed to various kinds of quantum hardware~\cite{peruzzo2014variational,kandala2017}. The VQE method is a hybrid quantum-classical approach in which energies are evaluated on quantum hardware or simulators, and classical computers perform the algorithm to optimize the variational parameters. The repeated evaluation of the quantum circuits can be costly, especially if there are slow convergences of the variational parameters. In the first stage of our calculations, we carried out noise-free VQE simulations to obtain optimal variational parameters. These simulations employed Qiskit's Aer simulator with the simultaneous perturbation stochastic approximation optimizer and EfficientSU2 two-level circuit as the ansatz~\cite{matthew_treinish_2022_6784303}. At this point, we limited ourselves to the 1 COVO basis and three internuclear distances to evaluate for the sake of computational cost. A batch of four circuits, shown in Figure \ref{fig:Circuit}, were evaluated to compute the energy. These circuits all require two qubits and consist of 8 $R_y$ and 8 $R_y$ rotation gates for the 16 variational parameters in the ansatz, 3 Controlled-X (or Controlled-NOT) gates, and 2 measurements on the two qubits, along with either 0, 1, or 2 Hadamard gates \cite{mcmahon2007quantum}. These results reproduced FCI energies to less than a milliHartree for the three geometries. Currently, proposals for robust quantum error correction require qubit numbers and performance that are not yet available via Cloud-based NISQ devices today~\cite{shor1995,cory1998,reed2012}, so before executing the circuits on the H1-1 quantum computer, we wanted to ensure that noise played a manageable role in computing the ground-state energies. So, for the second stage of the calculations, we used the optimal variational parameters obtained from the noise-free simulation with the noisy Quantinuum emulator that mimics the actual behavior of the Quantinuum H1-1 quantum computer. The error from noise was corrected using the post-processing mitigation technique called the full calibration measurement correction fitter, which measures a circuit with an expected result several times to construct a calibration matrix. The corresponding circuits can be seen in Figure \ref{fig:calibration}. There was a limit to the number of times a circuit could be executed, which was 500 times, so we performed 500 shots in the simulation. Between the circuits for the energy evaluation and the error mitigation, each complete run consumed nearly 80 credit units, a significant portion of our allocation. The number of credit units required is computed using the following formula:
\begin{eqnarray*}
    \text{Units} = 5 + C(N_{1q} + 10N_{2q}+ 5N_{m})/5000,
\end{eqnarray*}
where $C$ is the shot count, $N_{1q}$ is the number of one-qubit operations, $N_{2q}$ is the number of two-qubit operations, and $N_{m}$ is the number of state preparation and measurement operations per circuit. After convincing evidence that the error from the noise for this circuit can be well tempered, we performed the last stage of the calculations, where the same energy evaluation and error mitigation technique was performed for 500 runs on the Quantinuum H1-1 quantum computer. 

For the three points, energies obtained on the H1-1 quantum computer and simulations reproduced the FCI values to less than 11 milliHartree (6.9 kcal/mol) when corrected for noise. These errors are expected to improve with a larger number of circuit runs, which was not feasible then. Given the advertised high-fidelity of qubits for the H1-1 quantum computer, error mitigation played a small role in hardware calculations and simulations, reducing all energies by 1-3 milliHartree. Overall, our results are promising, especially considering that another study of LiH showed error mitigated results ranging from $\sim$10-60 milliHartree along the potential energy surface~\cite{Kandala2019}.

\begin{figure}[]
    \centering
    \begin{tikzpicture}[scale=1.4]
    \begin{axis}[my style,grid=major,minor tick num=1,xmin=1.0,xmax=7.5,ymin=-0.77,ymax=-0.57,legend pos=north east, legend style={nodes={scale=0.6}}, legend cell align={left}]
        \addplot [smooth,thin,black]    table [x=R(Li-H), y=FCI,  col sep=comma] {data/LiH-Quantum-plotvals.csv};
        \addplot [only marks, mark=triangle*,red]  table [x=R(Li-H), y=Simulation, col sep=comma] {data/LiH-Quantum-plotvals.csv};
        \addplot [only marks, mark=triangle,red]  table [x=R(Li-H), y=SimNoMit, col sep=comma] {data/LiH-Quantum-plotvals.csv};
        \addplot [only marks, mark=square*,blue]  table [x=R(Li-H), y=H1-1, col sep=comma] {data/LiH-Quantum-plotvals.csv};
        \addplot [only marks, mark=square,blue]  table [x=R(Li-H), y=H1-1NoMit, col sep=comma] {data/LiH-Quantum-plotvals.csv};
        \legend{PW FCI - 1 COVO, H1-1 simulator w/ error correction, H1-1 simulator w/o error correction, H1-1 quantum computer w/ error correction , H1-1 quantum computer w/o error correction}
    \end{axis}
    \end{tikzpicture}  
    \caption{VQE simulations and calculations on the Quantinuum H1-1 quantum computer. Red triangles correspond to energies obtained with the H1-1 quantum computer simulator, while the blue squares correspond to the energies obtained using the H1-1 quantum computer. 
    Open triangles and square are used to represent energies before error mitigation, while the filled shapes are the error corrected values. We note both the quantum computer and simulator results are very good, and error mitigation has very little effect on the overall results.
    The energies are plotted with the FCI potential energy curve, given by the solid black line. A simple-cubic supercell (L=15.0 \AA) was used.}
	\label{fig:lih-quantum}
\end{figure}

\begin{table}[]
   \centering
   \begin{filecontents*}{data/LiH-Quantum2.csv}
R(Li-H),FCI,SimNoMit,Simulation,H1-1NoMit,H1-1
1.7,-0.76044,-0.76434,-0.76547,-0.75506,-0.75786
,,(-0.00390),(-0.00503),(-0.00539),(-0.00258)
3,-0.70928,-0.70946,-0.71135,-0.69688,-0.69837
,,(-0.00018),(-0.00207),(-0.01241),(-0.01091)
7,-0.64801,-0.64623,-0.64747,-0.6433,-0.64433
,,(-0.00178),(-0.00053),(-0.0047),(-0.00367)
\end{filecontents*}
\csvreader[
   tabular=c|ccccc,
   table head=$R$(Li-H) & FCI & Simulator w/o & Simulator w/ & H1-1 w/o & H1-1 w/ \\
   (\AA) & & Mitigation & Mitigation & Mitigation & Mitigation \\
         & & (error)    & (error)    & (error)    & (error)    \\\hline,
   late after last line=\\ \hline 
   ]{data/LiH-Quantum2.csv}{}{\csvlinetotablerow}
   \caption{Total energies in Hartree for LiH using the 1 COVO basis set for VQE simulations and hardware calculations on the Quantinuum H1-1 quantum computer. A simple-cubic supercell (L=15.0 \AA) was used. Values in parenthesis are error relative to FCI.}
   \label{tab:LiHQuantum}
\end{table}

\section{Conclusion}
\label{sec:conclusions}

In summary, we have extended the COVOs method to periodic systems at the $\Gamma$ point using the recently developed Filon integration strategy~\cite{bylaska2020filon} for two-electron periodic integrals, in which the integration of the first Brillouin zone is automatically incorporated. We would also like to note that Fig. 1 in  reference~\cite{bylaska2020filon} illustrates how these integrations can be generalized to include explicit integrations over the first Brillouin zone by it up into patches (see \url{https://materialstheory.springeropen.com/articles/10.1186/s41313-020-00019-9/figures/1}).
For an (N+1)-state Hamiltonian, the method is based on optimizing the virtual orbitals to minimize a small select CI Hamiltonian (i.e., COVOs) that contains configurations containing all N filled RHF orbitals and the one virtual orbital to be optimized.  Subsequent virtual orbitals are optimized in the same way, but with the added constraint of being orthogonal to the filled orbitals and the previously optimized virtual orbitals.  
The method was applied to the simple, but non-trivial, LiH molecule in a periodic system, and we were able to obtain good agreement between the total energies from aperioidic and periodic plane-wave FCI calculations. Also, as shown in Fig.~\ref{fig:LiH-periodic-aperiodic-covos}, the shapes of the periodic COVOs are basically the same as what is found for the COVOs from aperiodic calculations, which indicates that this extended periodic COVOs method can reproduce the results by the aperiodic COVOs method in our previous work~\cite{bylaska2021quantum}. Subsequent calculations showed that the correlation energy converged steadily as more virtual orbitals were included in the calculation.  With 18 virtual orbitals the correlation energies were found to be converged to less than 1 kcal/mol. 

To test the validity of the periodic COVOs method on a NISQ device, we carried out VQE simulations on the H1-1 quantum computer and its simulator.  It was found that the energies obtained using the H1-1 quantum computer were able to reproduce the FCI values to less than 11 milliHartree (6.9 kcal/mol) with a modest number of 500 shots performed; slightly less when corrected for noise. These errors are expected to improve with a larger number of circuit runs. For both simulation and hardware calculations, it was found that error mitigation played a small role, only reducing the energies by 1-3 milliHartree. These results were promising, and open the door to running larger molecular and crystalline systems on NISQ devices in the near future.



\appendix
\section{}
\label{sec:appendix}
\begin{appendix}

Total energies from periodic and aperiodic plane-wave FCI calculations with 1, 4, 8, 12 and 18 COVOs are shown in Table ~\ref{tab:lihpwfci} and \ref{tab:lihpwfci-apd}.
\newpage
\begin{table}[h!]
   \centering
   \begin{filecontents*}{data/LiH-periodic.csv}
      R(Li-H),1 COVO,4 COVOs,8 COVOs,12 COVOs,18 COVOs
        1.30  ,	-0.74131  , -0.76074      ,	-0.76224  ,	-0.76317  ,	-0.76362
        1.40  ,	-0.75131  ,	-0.77049  ,	-0.77196  ,	-0.77287  ,	-0.77333
        1.50  ,	-0.75719  ,	-0.77602  ,	-0.77747  ,	-0.77835  ,	-0.77881
        1.60  ,	-0.75998  ,	-0.77838  ,	-0.77981  ,	-0.78066  ,	-0.78112
        1.70  ,	-0.76044  ,	-0.77834  ,	-0.77976  ,	-0.78059  ,	-0.78103
        1.80  ,	-0.75918  ,	-0.77654  ,	-0.77795  ,	-0.77875  ,	-0.77919
        1.90  ,	-0.75668  ,	-0.77348  ,	-0.77489  ,	-0.77565  ,	-0.77609
        2.00  ,	-0.75331  ,	-0.76956  ,	-0.77097  ,	-0.77171  ,	-0.77215
        2.50  ,	-0.73123  ,	-0.74492  ,	-0.74652  ,	-0.74714  ,	-0.74759
        3.00  ,	-0.70928  ,	-0.72091  ,	-0.72319  ,	-0.72368  ,	-0.72430
        3.50  ,	-0.69108  ,	-0.70100  ,	-0.70474  ,	-0.70513  ,	-0.70618
        4.00  ,	-0.67684  ,	-0.68530  ,	-0.69151  ,	-0.69358  ,	-0.69377
        4.50  ,	-0.66613  ,	-0.67458  ,	-0.68316  ,	-0.68628  ,	-0.68666
        5.00  ,	-0.65837  ,	-0.67791  ,	-0.67902  ,	-0.68212  ,	-0.68283
        6.00  ,	-0.65061  ,	-0.67364  ,	-0.67703  ,	-0.67705  ,	-0.67848
        7.00  ,	-0.64801  ,	-0.67035  ,	-0.67340  ,	-0.67342  ,	-0.67461
        8.00  ,	-0.64800  ,	-0.67035  ,	-0.67340  ,	-0.67342  ,	-0.67461
        9.00  ,	-0.65061  ,	-0.67364  ,	-0.67703  ,	-0.67705  ,	-0.67848
        10.00 ,	-0.65837  ,	-0.67792  ,	-0.67903  ,	-0.68212  ,	-0.68284
        10.50 ,	-0.66613  ,	-0.67454  ,	-0.68315  ,	-0.68628  ,	-0.68667
        11.00 ,	-0.67684  ,	-0.68530  ,	-0.69151  ,	-0.69358  ,	-0.69377
        11.50 ,	-0.69108  ,	-0.70100  ,	-0.70474  ,	-0.70513  ,	-0.70618
        12.00 ,	-0.70928  ,	-0.72092  ,	-0.72319  ,	-0.72368  ,	-0.72430
        12.50 ,	-0.73123  ,	-0.74492  ,	-0.74652  ,	-0.74713  ,	-0.74759
        13.00 ,	-0.75331  ,	-0.76956  ,	-0.77097  ,	-0.77171  ,	-0.77214
        13.10 ,	-0.75668  ,	-0.77348  ,	-0.77489  ,	-0.77565  ,	-0.77609
        13.20 ,	-0.75918  ,	-0.77654  ,	-0.77795  ,	-0.77874  ,	-0.77919
        13.30 ,	-0.76044  ,	-0.77834  ,	-0.77976  ,	-0.78058  ,	-0.78103
        13.40 ,	-0.75998  ,	-0.77838  ,	-0.77981  ,	-0.78066  ,	-0.78112
        13.50 ,	-0.75719  ,	-0.77602  ,	-0.77747  ,	-0.77835  ,	-0.77881
        13.60 ,	-0.75131  ,	-0.77049  ,	-0.77196  ,	-0.77287  ,	-0.77333
        13.70 ,	-0.74131  ,	-0.76074  ,	-0.76224  ,	-0.76317  ,	-0.76362
\end{filecontents*}
\csvreader[
   tabular=c|ccccc,
   table head= & PW FCI & PW FCI & PW FCI & PW FCI & PW FCI\\
    $R$(Li-H) & 1 COVO & 4 COVOs & 8 COVOs & 12 COVOs & 18 COVOs\\
   (\AA) & (Hartree) & (Hartree) & (Hartree) & (Hartree) & (Hartree)\\\hline,
   late after last line=\\ \hline 
   ]{data/LiH-periodic.csv}{}{\csvlinetotablerow}
   \caption{Total energies as a function of distance for the LiH molecule from periodic plane-wave FCI calculations with 1, 4, 8, 12, and 18 COVOs. A simple-cubic supercell (L=15.0 \AA) was used.}
   \label{tab:lihpwfci}
\end{table}

\begin{table}[h!]
   \centering
   \begin{filecontents*}{data/LiH-aperiodic.csv}
R(Li-H),1 COVO,4 COVOs,8 COVOs,12 COVOs,18 COVOs
1.30, -0.74087,-0.76030,-0.76181,-0.76273,-0.76319
1.40, -0.75086,-0.77002,-0.77150,-0.77240,-0.77286
1.50, -0.75671,-0.77552,-0.77697,-0.77785,-0.77831
1.60, -0.75947,-0.77784,-0.77928,-0.78013,-0.78058
1.70, -0.75990,-0.77777,-0.77919,-0.78002,-0.78046
1.80, -0.75860,-0.77593,-0.77734,-0.77813,-0.77858
1.90, -0.75606,-0.77283,-0.77424,-0.77500,-0.77544
2.00, -0.75266,-0.76885,-0.77027,-0.77100,-0.77144
2.50, -0.73040,-0.74397,-0.74559,-0.74620,-0.74666
3.00, -0.70839,-0.71976,-0.72215,-0.72262,-0.72327
3.50, -0.69047,-0.69983,-0.70404,-0.70443,-0.70556
4.00, -0.67722,-0.68469,-0.69232,-0.69454,-0.69472
4.50, -0.66867,-0.68639,-0.68766,-0.69015,-0.69077
5.00, -0.66424,-0.68613,-0.68777,-0.68922,-0.69032
6.00, -0.66366,-0.68713,-0.68942,-0.68943,-0.69025
7.00, -0.66372,-0.68739,-0.68945,-0.68946,-0.69011
\end{filecontents*}
\csvreader[
   tabular=c|ccccc,
   table head= & PW FCI & PW FCI & PW FCI & PW FCI & PW FCI\\
   $R$(Li-H) & 1 COVO & 4 COVOs & 8 COVOs & 12 COVOs & 18 COVOs\\
   (\AA) & (Hartree) & (Hartree) & (Hartree) & (Hartree) & (Hartree)\\\hline,
   late after last line=\\ \hline 
   ]{data/LiH-aperiodic.csv}{}{\csvlinetotablerow}
   \caption{Total energies in Hartree as a function of distance for the LiH molecule from aperiodic plane-wave FCI calculations with 1, 4, 8, 12, and 18 COVOs. A simple-cubic supercell (L=15.0 \AA) was used.}
   \label{tab:lihpwfci-apd}
\end{table}
\end{appendix}

\begin{backmatter}

\section*{Acknowledgements}
This material is based upon work supported by the U.S. Department of Energy (DOE), Office of Science, Office of Basic Energy Sciences, Chemical Sciences, Geosciences, and Biosciences (CSGB) Division through its ``Embedding Quantum Computing into Many-body Frameworks for Strongly Correlated  Molecular and Materials Systems'' project  at Pacific Northwest National Laboratory (PNNL). This work was also supported by the Quantum Science Center (QSC), a National Quantum Information Science Research Center of the U.S. Department of Energy (DOE). KMR also acknowledges support from BES CSGB's Geosciences program at PNNL.
We also would like to thank the DOE BES Chem CCS, DOE BES Geosciences and DOE Advanced Scientific Computing Research (ASCR) ECP NWChemEx (17-SC-20-SC) programs for their support of software development for high-performance computers and computer time needed to carry out the work. PNNL is operated for the U.S. Department of Energy by the Battelle Memorial Institute under Contract DE-AC06-76RLO-1830.  
This research used computational resources at the Pacific Northwest National Laboratory and resources of the National Energy Research Scientific Computing Center (NERSC), a User Facility supported by the Office of Science of the U.S. DOE under Contract No. DE-AC02-05CH11231, and the Argonne ALCF computing center through their early science program. We would also like to thank the NWChem project team and the people that have helped the progress of the NWChem software over the years.  
Lastly, we would like to thank the Azure Quantum Credits Program for a generous grant of computer time to run on the on the Quantinuum’s ion trap quantum computers.



\section*{Availability of data and materials}
The codes developed are available in the NWChem software (\url{https://github.com/nwchemgit/nwchem}, or in near future NWChemEx~\cite{kowalski2021nwchem}~\url{https://github.com/ebylaska/PWDFT}). Input and output decks used to test the codes are available from the corresponding author. The data sets containing the one and two electron integrals based on COVOs that were used for the VQE quantum computing calculations can be found in the following github repository,
\url{https://github.com/duosong/PWLiH-Data}.

\section*{Authors’ contributions}
All authors contributed to the writing of the article. DS and EJB developed the COVOs model and codes. DS, NB, GP, CG and EJB contributed to application of the codes. NB, GP, BP, CG, GL and MR contributed to the error mitigation calculations.  KK, KMR, and MR were the PIs of computing projects used in the development of the codes and quantum software libraries, and  MR was also in charge of the computing project that provided access to the Quantinuum quantum computer.  All authors read and approved the final manuscript.

\section*{Competing interests}
The authors declare that they have no competing interests.

\end{backmatter}

\bibliographystyle{bmc-mathphys}
\bibliography{Manuscript}


\end{document}